\documentclass{aastex63}

\newcommand{\RomanNumeralCaps}[1]
    {\MakeUppercase{\romannumeral #1}}

\submitjournal{\apj}

\shorttitle{Neutron star X-ray binaries associated with supernova remnants}
\shortauthors{}

\begin{document}

\title{POPULATION SYNTHESIS OF NEUTRON STAR X-RAY BINARIES ASSOCIATED WITH SUPERNOVA REMNANTS}

\author{Ze-Pei Xing}
\affil{Department of Astronomy, Nanjing University, Nanjing 210023, China;
lixd@nju.edu.cn}
\affil {Geneva Observatory, University of Geneva, Chemin des Maillettes 51, 1290 Sauverny, Switzerland;
Zepei.Xing@unige.ch}
\author{Xiang-Dong Li}
\affil{Department of Astronomy, Nanjing University, Nanjing 210023, China;
lixd@nju.edu.cn}
\affil{Key Laboratory of Modern Astronomy and Astrophysics, Nanjing University,
Ministry of Education, Nanjing 210023, China}

\begin{abstract}

Neutron star X-ray binaries (NS XRBs) associated with supernova remnants (SNRs) are youngest X-ray binaries that can provide insights into early evolution of X-ray binaries and formation properties of neutron stars. There are an increasing number of NS XRBs discovered to be harbored in SNRs in our and nearby galaxies. In this work, we perform binary population synthesis calculations to simulate the population of NS XRBs associated with SNRs for different types of companions, including Roche-lobe overfilling main-sequence stars, Be stars and supergiants. We estimate their birthrates and present the distributions of orbital parameters and companion mass for each type of companions. Our calculations show that the majority of them are Be X-ray binaries (BeXRBs) and that a few BeXRBs are expected to be associated with SNRs in a Milky Way-type galaxy.     
\end{abstract}



\section{Introduction} \label{sec:intro}

Neutron star X-ray binaries (NS XRBs) within supernova remnants (SNRs) are very young systems since the lifetime of a SNR is typically $\lesssim 10^{5}\ \rm{yr}$, which is much less than the evolutionary time of XRBs. These systems can provide valuable information on the early evolution of XRBs as well as the properties of newborn NSs such as their spins and magnetic fields. The short lifetime of SNRs also results in the rareness of such systems. To date, we have only several confirmed cases in our and nearby galaxies \citep[][for a review]{Li20}. Two of them are Be X-ray binaries (BeXRBs) in the Small Magellanic Cloud, i.e., SXP 1062 \citep{HB12,Ha12} and SXP 1323 \citep{Gv19}. In the Large Magellanic Cloud, DEM L241 \citep{Se12} probably hosts an accreting NS \citep{vanS19} and an O-type companion star. The recently discovered SNR, MCSNR J0513-6724, hosts a supergiant X-ray binary (SGXB) \citep{Ma19}. For the Galactic NS XRB Cir X-1, which is associated with the SNR G322.1+0.0 \citep{He13}, the nature of its companion is still under debate. Most recently, \citet{Ma21} reported a new BeXRB, XMMU J050722.1-684758, possibly associated with MCSNR J0507-6847 in the Large Magellanic Cloud. In Table \ref{tab:1}, we summarize the observational and derived properties of these XRBs and the associated SNRs.  

SXP 1062 has a long-period X-ray pulsar with a spin period of $1062\ \rm{s}$ \citep{HB12}. According to the multi-wavelength study carried out by \citet{GG18}, the orbital period of SXP 1062 is $ 668\pm10\ \rm{d}$, which is compatible with former studies \citep{Sch12,St13}, and its eccentricity is between $0.4$ and $0.88$. Two individual groups have obtained similar estimates of the kinematic age of the associated SNR MCSNR J0127-7332, which are $20-40\ \rm{kyr}$ \citep{HB12} and 16 kyr \citep{Ha12}, respectively. SXP 1323 also shows long-period pulsations with a period of $1323\ \rm{s}$ \citep{HP05}. However, the orbital period is only $26.2\ \rm{d}$ \citep{SC06,Ca17}, meaning that it does not follow the empirical relationship between the orbital and pulse periods of BeXRBs. \citet{Gv19} analyzed the information of its associated SNR and obtained an age of $\sim 40\ \rm{kyr}$. Both NSs in these two young BeXRBs show long spin periods, implying that NSs can spin down significantly within tens of thousands of years. And their discoveries indicate that the formation timescales of HMXBs can be shorter than the typical lifespan of a SNR, while it was traditionally expected that the accretion process would begin after the SNR has already faded \citep{Bv91}. Through detailed timing and spectral analysis of the BeXRB XMMU J050722.1-684758, \citet{Ma21} found pulsations with a period of $570\ \rm{s}$ and an orbital period of $40.2\ \rm{d}$. The BeXRB is located near the geometric center of MCSNR J0507-6847, which is about $43-63\ \rm{kyr}$ old. Thus, it could be a new discovered BeXRB associated with a SNR.

\citet{Se12} found that a point-like source CXOU 053600.0-673507 in the SNR DEM L241 has a X-ray luminosity $\sim 2\times10^{35} \rm{erg}\ \rm{s^{-1}}$. Based on the variability in X-rays and the fact that its optical counterpart is an O5\RomanNumeralCaps{3}(f) star, they suggested that it is likely a HMXB with a NS accreting material from an O-star's wind. And they estimated the age of the SNR to be $>(50-70)\ \rm{kyr}$. Later, \cite{Cob16} discovered luminous gamma-ray emission which modulated at a period of $10.3\ \rm{d}$. Then, from optical spectroscopic observations, \citet{vanS19} obtained an eccentricity of $0.40\pm0.07$ and the mass function $f = 0.0010\pm0.0004\ M_{\odot}$, which favors a pulsar wind driven but not accretion driven system. \citet{Ma19} reported the discovery of a high-mass X-ray binary (HMXB) with a B2.5Ib supergiant companion associated with the supernova remnant MCSNR J0513-6724. Analyzing the optical light curve from Optical Gravitational Lensing Experiment (OGLE), they obtained an orbital period of $2.2325\pm0.0003\ \rm{d}$. And the X-ray timing analysis revealed a possible spin period of $4.399\ \rm{s}$ for the NS. From the temperature and the ionization timescale of the SNR, they estimated the age to be $< 6\ \rm{kyr}$, indicating that it may be the youngest known HMXB. Moreover, assuming that the NS is at the spin equilibrium stage, they estimated the upper limit of the NS's magnetic field to be $< 5 \times 10^{11}\ \rm{G}$, which is pretty low for a young NS. But \citet{Ho20} argued that the NS may be in the propeller phase and found a relatively high magnetic field of $>$ a few $\times 10^{13}\ \rm{G}$.

Cir X-1 is a confirmed NS XRB associated with a SNR in the Milky Way, whose age is less than $4600\ \rm{yr}$ \citep{He13}. It has an orbital period of $16.6\ \rm{d}$ \citep{Ka76,Wh77} and a highly variable X-ray luminosity that can reach $>10^{38}\ \rm{erg}\ \rm{s}^{-1}$ \citep{Li10,He15}. The identification of its compact star as a NS resulted from the observed Type I X-ray bursts from it \citep{Te86,Pa10,Li10}. Type I X-ray bursts usually occur in relatively old low-mass X-ray binaries in which the NS has a weak magnetic field $\ll 10^{12}\ \rm{G}$ \citep{Fu81,Bi98}. As a result, Cir X-1 was originally regarded as a low-mass X-ray binary. More recently, \citet{Jon07} reported that the detected Paschen absorption lines fit a supergiant of spectral type of B5-A0, but they could not rule out the possibility that the spectrum comes from the accretion disk. They also got a moderate eccentricity of $\sim 0.45$ for the system. Later, \citet{Joh16} proposed that Cir X-1 may have a low-temperature and underdense companion star filling its Roche lobe at periastron in an orbit with an eccentricity of $\sim 0.4$. \citet{Sch20} found that the strongest lines in the periastron observations show a blueshift of $\sim 400\ \rm{km}$, indicating an ionized wind from a B5Ia supergiant. But they also pointed out that Cir X-1 showed two patterns of outbursts related to BeXRBs.

These young XRBs provide unique insights on the formation and early evolution of NS XRBs. It is thrilling to have more discoveries and observations of these sources for further studies. Meanwhile, it is essential to understand their formation in theory. In this paper, we carry out a population synthesis study of the NS XRBs within SNRs. The purpose of this work is two folds. On one hand, we try to predict the distributions of the binary quantities, which may be useful for further observations. On the other hand, population synthesis calculations can provide some clues to the properties of observed sources whose nature is poorly determined. In section \ref{sec:method}, we introduce the binary population synthesis (BPS) method that we use to generate the primordial binaries. In section \ref{sec:res}, we present the results from the BPS calculations. We present discussions in section \ref{sec:dis} and make a brief summary in section \ref{sec:sum}.

\begin{deluxetable}{cccccc}
\tablecaption{Known XRBs and Candidates Associated with SNRs\label{tab:1}}


\tablehead{\colhead{XRB}& \colhead{SNR} & \colhead{Orbital period} & \colhead{Eccentricity}  & \colhead{Spin period} &\colhead{Age} \\ 
\colhead{}& \colhead{} &\colhead{(d)} & \colhead{}  & \colhead{(s)}& \colhead{(kyr)}  } 

\startdata
SXP 1062 (1) & SNR J0127.7-7333& 668$\pm$10 & 0.4-0.88  & 1062&$\sim$ 20-40   \\
SXP 1323 (2) & - & 26.2 &  - & 1323 &$\sim$40  \\
XMMU J050722.1-684758 (3) & MCSNR J0507-6847 & 40.2 & - &570&$\sim$ 43-63 \\
LXP 4.4 (4) &MCSNR J0513-6724 &2.2324$\pm$0.0003 & -  &4.4& $<$6   \\
CXOU 053600.0-673507 (5)& DEML241 & 10.301 &  0.40$\pm$0.07  &-& $>$50-70   \\
Cir X-1 (6) & G322.1+0.0 & 16.6 &  0.4-0.45  &-& $<$4.6  \\
\enddata
\tablerefs{(1) \citet{Ha12,HB12,GG18} (2) \citet{HP05,Gv19} (3) \citet{Ma21} (4) \citet{Ma19} (5) \citet{Se12,vanS19} (6) \citet{He13}}

\end{deluxetable}

\section{Generating Primordial Binaries} \label{sec:method}

We use the BPS code BSE that was initially developed by \citet{H02} incorporating some modifications made by \citet{Y14}. With BSE we can generate a large number of primordial binaries with various initial parameters. We evolve the binaries until an NS is born in a binary and then focus on the newborn XRBs. We will introduce some critical treatments of physical processes of binary modelling employed in the code.

As the primary star evolves to fill its Roche lobe in the first place, it will transfer mass to the secondary through Roche-lobe overflow (RLOF). The mas transfer rate and the accretion rate are crucial for the evolution of the binary since rapid mass accretion can cause significant expansion of the secondary, which may lead to a contact phase \citep{NE01} and further to a common envelope (CE) evolution phase. BPS studies have shown that the rotation-dependent mass-transfer model can reproduce some kinds of Galactic binaries such as Be-BH systems \citep{Y14}, Wolf Rayet-O systems \citep{Y16}, and double NS systems \citep{Y18}. Thus, we assume that the accretion rate is related to the rotation of the secondary and equals to the mass transfer rate multiplied by $(1-\frac{\omega}{\omega_{\rm{cr}}})$ \citep{Y14}, where $\omega$ is the secondary's angular velocity and $\omega_{\rm{cr}}$ is a critical value above which the accretion process would cease. The rotation-related accretion modification was first proposed by \citet{S09}. They suggested that for a fast-rotating star, its surface layers are less gravitationally bound and a star with a rotation velocity close to the critical value would not accrete more material. This inefficient accretion model makes the mass ratio of the primary to the secondary for stable mass transfer reach as high as $\sim 6$, meaning that a large number of the primordial binaries can avoid the CE process. If the binary enters CE evolution, the frictional torque of the secondary converts the orbital energy into internal energy of the primary's envelope and transfers the angular momentum of the secondary to the envelope, leading to a significant orbital decay. If the secondary can successfully expel the envelope and avoid the merger, it will leave a binary in a close orbit consisiting of the core of the primary star and the secondary. To treat the inspiral process, we employ the $\alpha_{\rm{CE}}$ prescription and adopt the calculations of \citet{Xu10} for the binding energy parameter $\lambda$. Generally $0<\alpha_{\rm{CE}}\leq 1$ \citep{I13}, and we take $\alpha_{\rm{CE}}$ to be $1.0$ in our reference model. 

Finally, the primary star explodes to form a NS. We consider two types of SN explosions, core-collapse SNe (CCS) and electron-capture SNe (ECS), that can give birth to NSs. From \citet{H02}, the helium core mass at the asymptotic giant branch base determines the mass of the CO core formed later, which is related to the different types of explosions. Specifically, if the helium core mass is lower than $1.83 M_{\odot}$, a degenerate CO core will form subsequently, leading to the formation of a CO white dwarf. If the helium core mass is higher than $2.25 M_{\odot}$, the star will form a non-degenerate CO core which will eventually lead to CCS. Between these two critical masses, the star may form a partially degenerate CO core. And if the core reaches $1.08 M_{\odot}$, it will burn into a degenerate ONe core without an explosion. Furthermore, if such an ONe core is more massive than $1.38 M_{\odot}$, it will become a NS through ECS. During the SN explosion, the NS will receive a natal kick that can cause an eccentric orbit or disruption of the system. By analyzing the proper motions of pulsars, \citet{Ho05} found that the kick velocity follows a Maxwellian distribution with a dispersion velocity of $\sigma_{k} = 265\ \rm{km}\ \rm{s}^{-1}$. Later, \citet{V17} proposed that a bimodal distribution with $\sigma_{k} = 80$ and $320\ \rm{km}\ \rm{s}^{-1}$ fits the birth velocity distribution of pulsars better, which may result from the two types of SNe mentioned above. The kick velocity can be the key factor affecting the birthrate and the initial orbital properties of newborn XRBs, so we adopt different dispersion velocities in the calculations to investigate how they can affect the formation of young XRBs. For ECS, we adopt $\sigma_{\rm{k,ECS}} = 40\ \rm{km}\ \rm{s}^{-1}$ \citep{D06}; for CCS, we adopt $\sigma_{\rm{k,CCS}}= 150$ and $300\ \rm{km}\ \rm{s}^{-1}$. 

\begin{figure}
\plotone{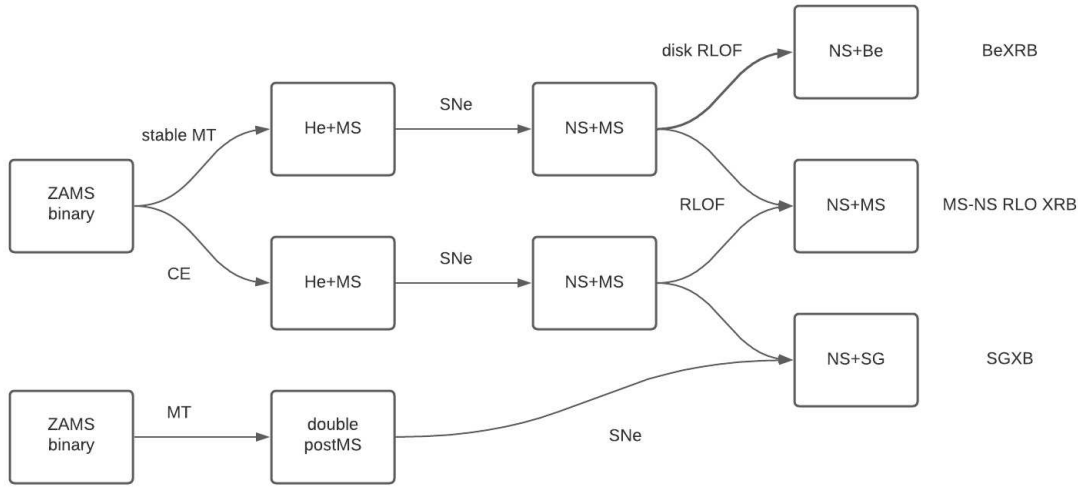}
\caption{Flowchart showing the formation channels of the three types of XRBs. MT means mass transfer, MS is main sequence, and He means helium star. \label{fig:1}}
\end{figure}

We consider three types of companions: Roche-lobe overfilling main-sequence stars, Be stars ,and supergiants. Figure \ref{fig:1} is a flowchart showing the formation channels of these three types of XRBs that we take account of in this work. For NS-MS Roche-lobe overfilling X-ray binaries, the primordial binary either experienced stable mass transfer or survived CE evolution. After the primary evolved to form a NS, we regard them as XRBs if the secondary overfilled the Roche lobe and initiated the mass transfer when it was still on the main sequence. For the formation of BeXRBs, we exclude the ones that underwent CE evolution. If the decretion disk RLOF happens in a NS-Be system, it is regarded as a BeXRB. And for SGXBs consisting of a NS and a supergiant companion, they have two possible formation channels. One is that, after the primary became a NS, the secondary left the main sequence within $10^{5}\ \rm{yr}$. The other is that, the initial mass ratio is close to unity, so the binary evolved to be a double post-main sequence binary and then the primary exploded to be a NS. The two stars might have undergone stable mass transfer or CE evolution. The secondary evolved to be a SG before the SNe. And when the primary exploded to be a NS, the binary may become a SGXB. We will have more detailed discussions on each of the binaries in the following section.

For each calculation, we generate $10^{7}$ primordial binary systems. We assume that the primary mass follows the initial mass function suggested by \citet{K93} and the mass ratio of the secondary to the primary has a uniform distribution between 0 and 1. The binary separation distribution is logarithmically flat between the limits of 3 and $10^{4}\ R_{\odot}$. For the star formation rate (SFR), we take a moderate value of $3\ M_{\odot}\ \rm{yr}^{-1}$ as a default for Milky-Way like Galaxies over a period of $10\ \rm{Gyr}$. Furthermore, we consider variations in the input parameters and investigate their influence on the results. For different types of donor stars, we adopt different sets of parameters to perform parameter studies.

\section{Results} \label{sec:res}

\subsection{NS-MS Roche-lobe overfilling X-ray binaries}       
\begin{figure}
\gridline{\fig{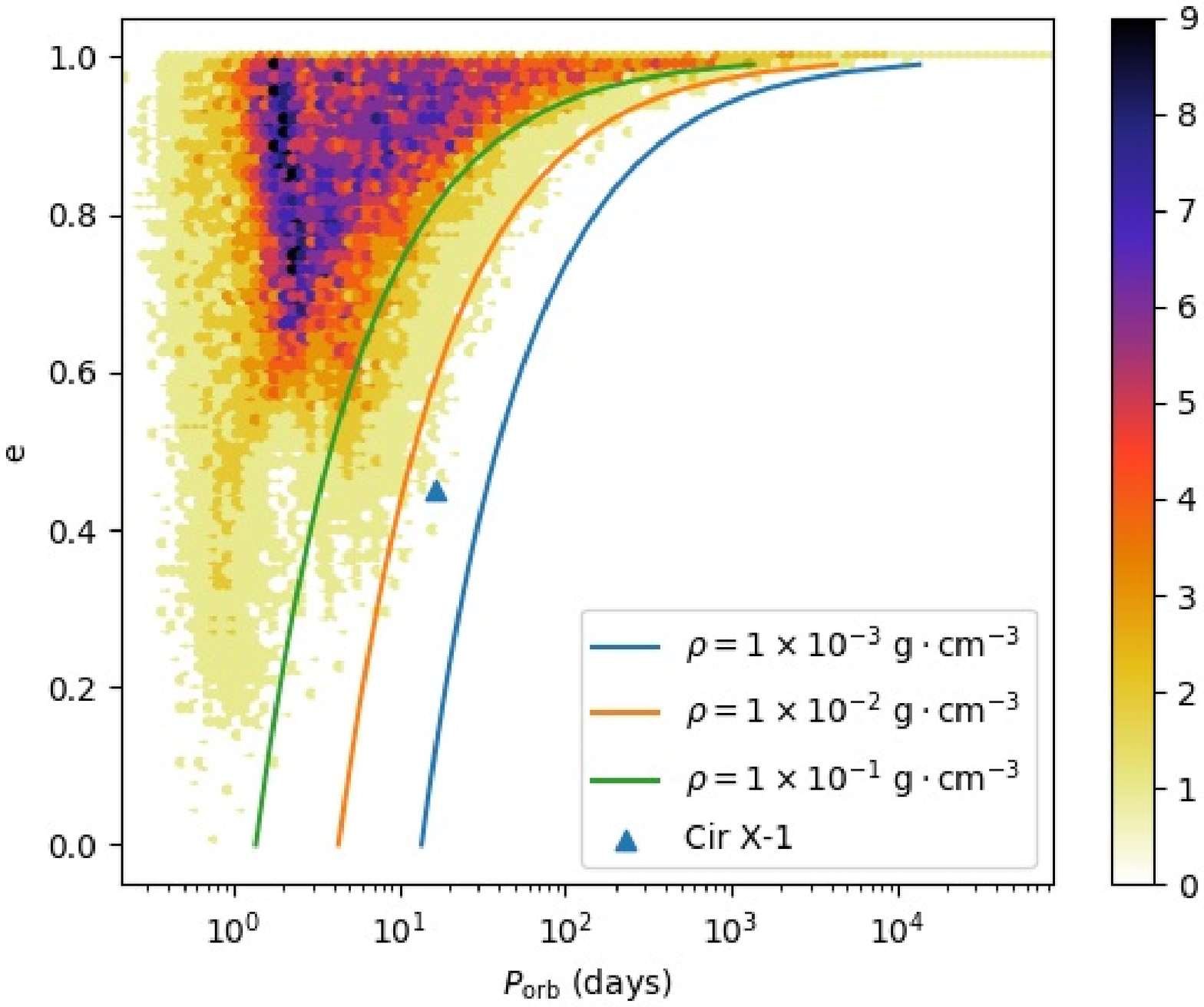}{0.4\textwidth}{}
          \fig{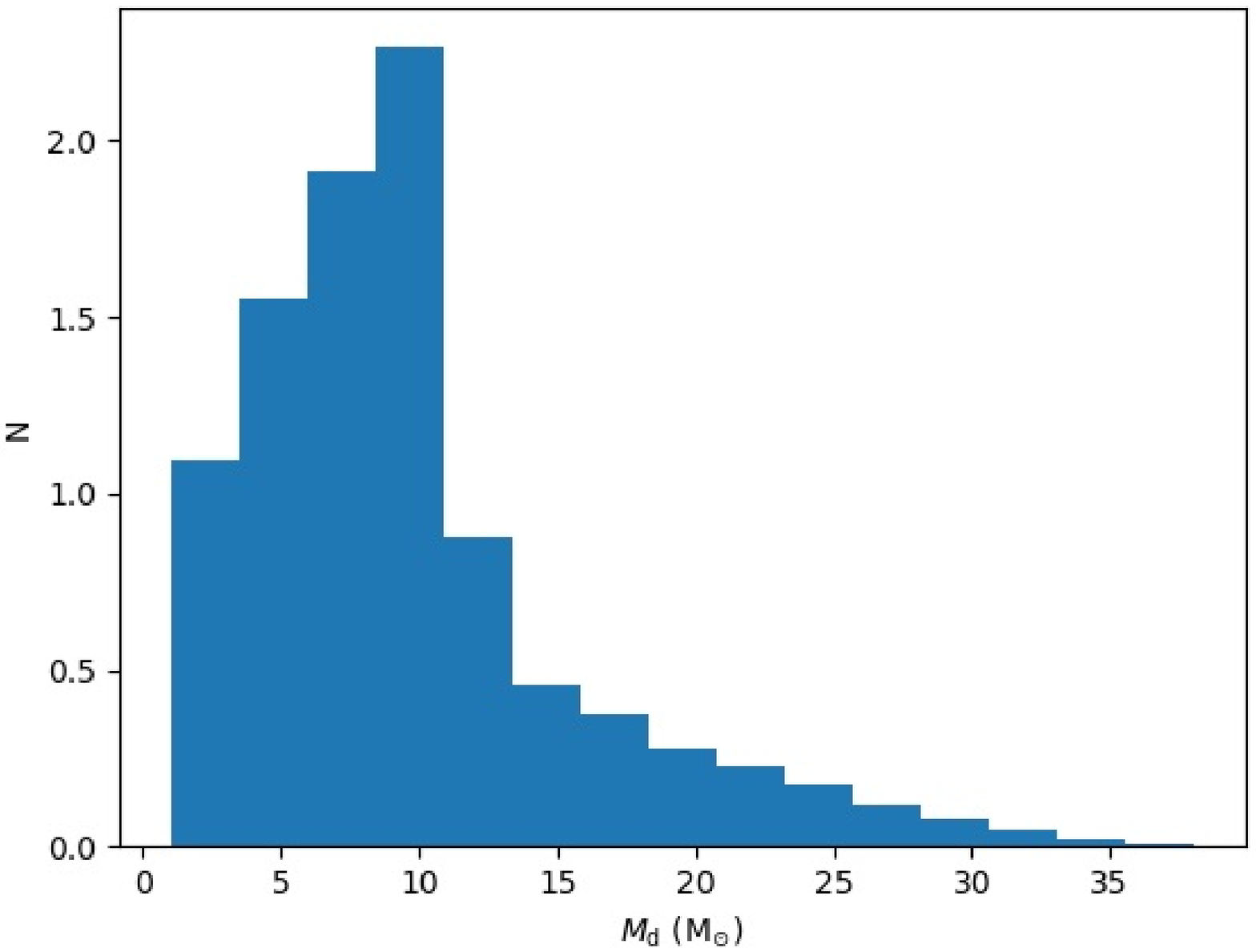}{0.4\textwidth}{}
          }
\gridline{\fig{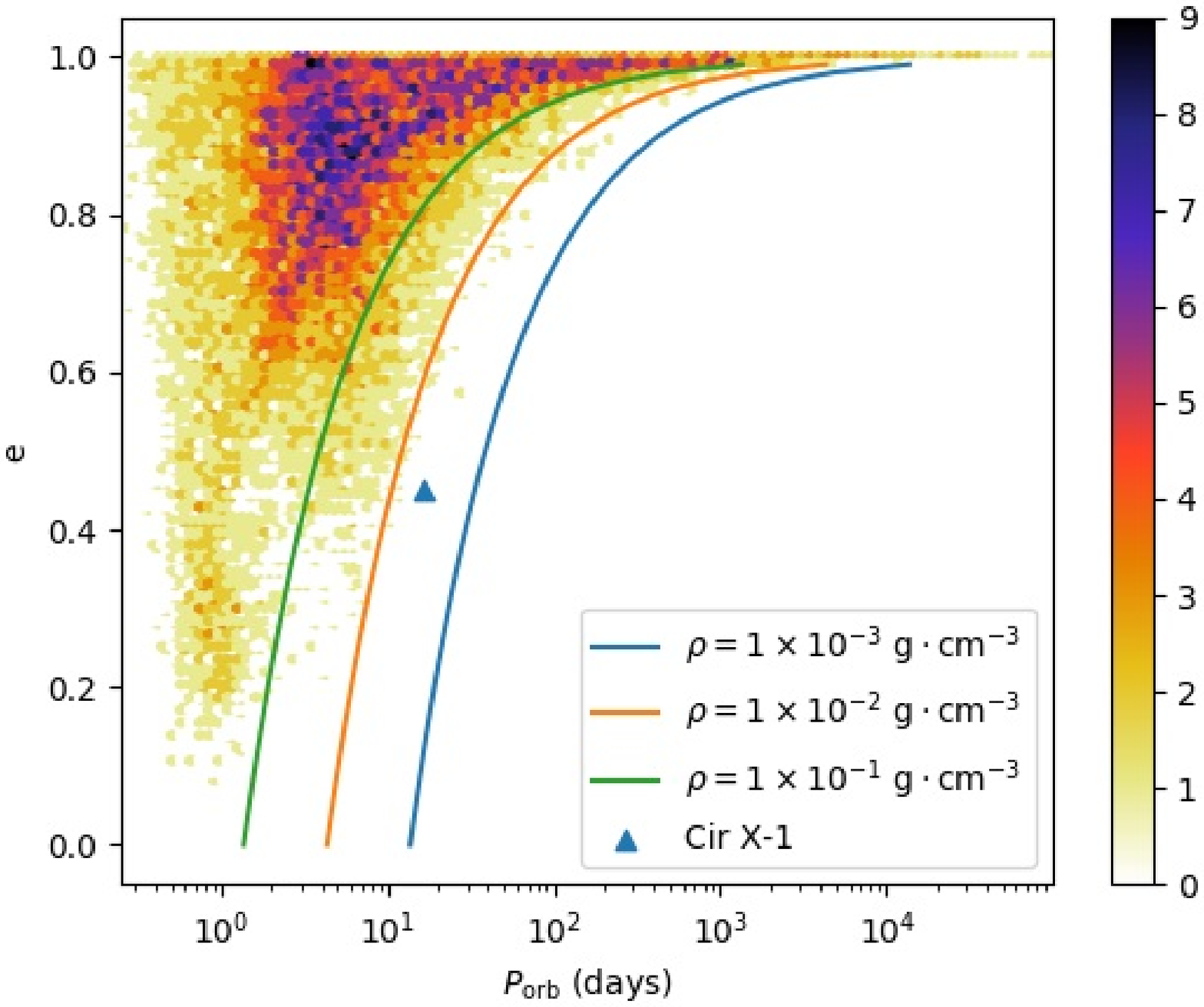}{0.4\textwidth}{}
          \fig{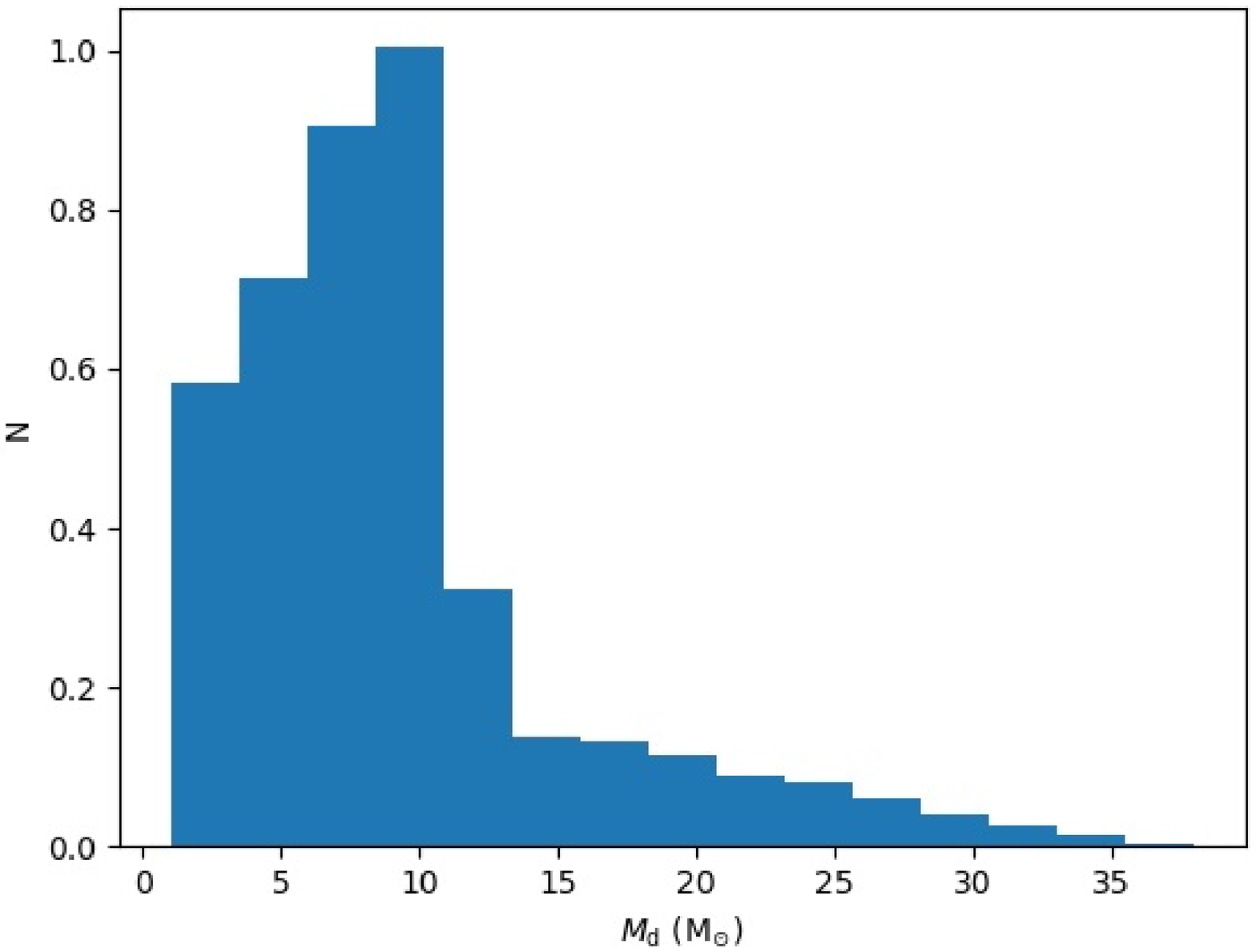}{0.4\textwidth}{}
          }
\caption{The distributions of the orbital parameters and the donor mass for NS-MS Roche-lobe overfilling XRBs within $10^{5}\ \rm{yr}$ after the SNe. In the left panel, the three lines represent the isodensity lines for stars that just overfill the Roche lobe at periastron. The blue triangle shows the position of Cir X-1. The top and bottom panels correspond to $\sigma_{\rm{k,CCS}}= 150$ and $300\ \rm{km}\ \rm{s}^{-1}$, respectively.\label{fig:2}}
\end{figure}

We first consider the possibility that the companion is a main-sequence star that fills its Roche lobe at periastron quickly after the birth of the NS. In BPS codes, no systematic treatments of mass transfer of XRBs in an eccentric orbit have been employed. We calculate the Roche lobe radius of a star at the periastron of an eccentric orbit using the approximated form \citep{E83,Se07}:
\begin{equation}
R_{{\rm L,peri}}=a(1-e)\frac{0.49q^{2/3}}{0.6q^{2/3}+\ln{(1+q^{1/3})}},
\end{equation}
where a is the semimajor axis, e is the eccentricity, and $q=M_{{\rm d}}/M_{{\rm NS}}$. In the calculation, we first select the binaries in which the companion fills its Roche lobe at periastron within $\sim 10^{5}\ \rm{yr}$, which is the typical observable time limit of SNRs, after the SN that forms a NS. Since the companions are in their main-sequence stage, it is extremely unlikely for a companion star that initially does not fill its Roche lobe at periastron to evolve and overflow its Roche lobe within $\sim 10^{5}\ \rm{yr}$. So, most of the selected binaries start transferring mass immediately after the NSs form. Figure \ref{fig:2} shows the distributions of these binaries in the eccentricity versus orbital period plane (the left panel) and the distribution of the companion mass (the right panel). The top and bottom panels correspond to the dispersion velocities for core-collapse supernova of $150\ \rm{km}\ \rm{s}^{-1}$ and $300\ \rm{km}\ \rm{s}^{-1}$, respectively. In the $P_{\rm{orb}}-e$ panel, we use different colors to denote the relative numbers of the sources in a specific hexagonal bin (the actual numbers should be multiplied by $10$). And this applies to the following similar plots. From the left panel, we can see that most of the binaries have relatively high eccentricities $(> 0.6)$ and those with eccentricities close to unity can reach long orbital periods ($\gtrsim 10^{4}$ days). The blue triangle shows the position of Cir X-1. In the right panel, the companion mass distribution depends on the mass distribution of the primary. And in a large fraction of the binaries, the masses of the companion stars are $\lesssim 10-12\ M_{\odot}$. The estimated birthrates with $\sigma_{\rm{k,CCS}}= 150$ and $300\ \rm{km}\ \rm{s}^{-1}$ are $9.47\times 10^{-5}$ and $4.23\times 10^{-5}\ \rm{yr}^{-1}$, respectively.         

\begin{figure}
\gridline{\fig{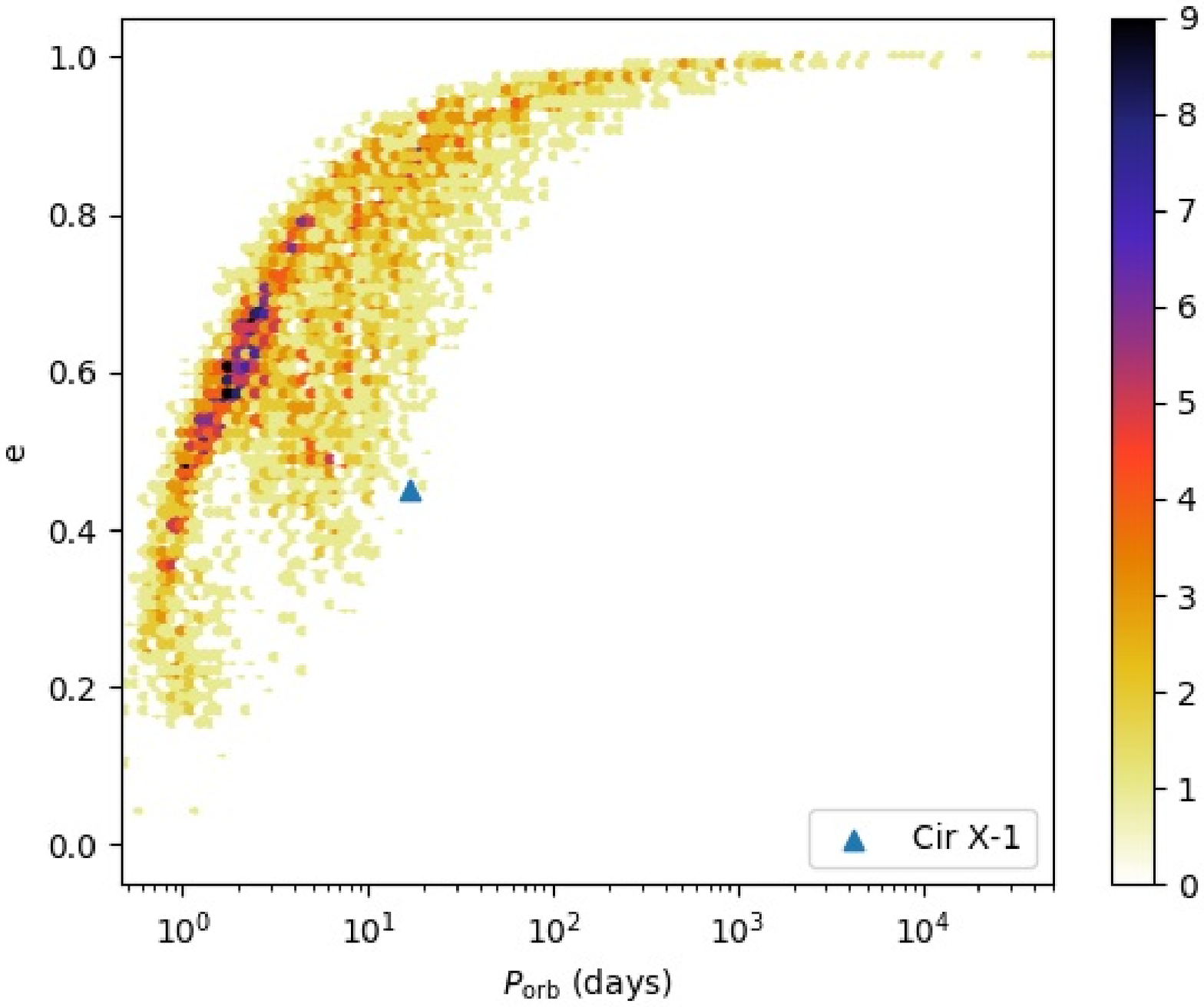}{0.4\textwidth}{}
          \fig{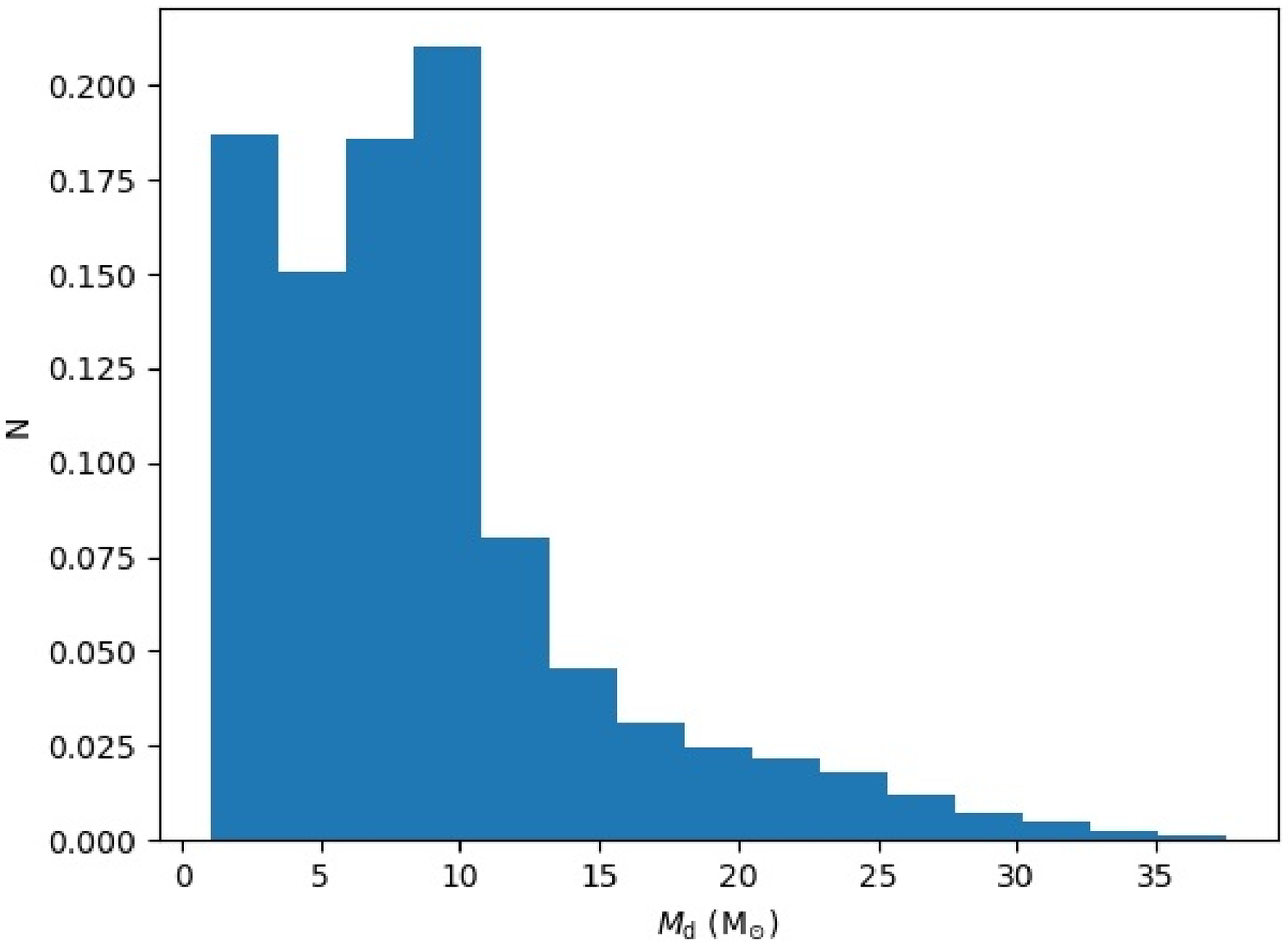}{0.4\textwidth}{}
          }
\gridline{\fig{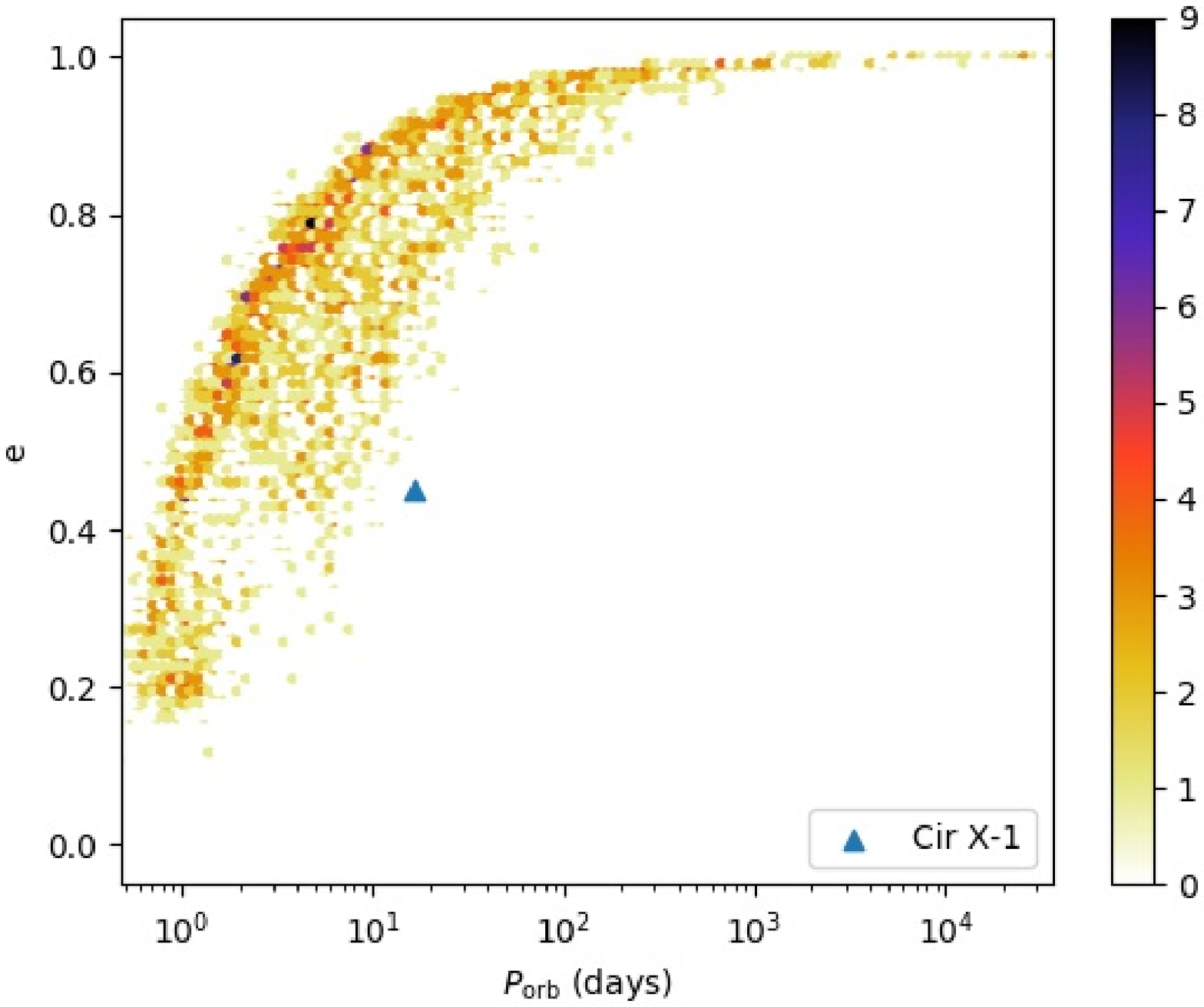}{0.4\textwidth}{}
          \fig{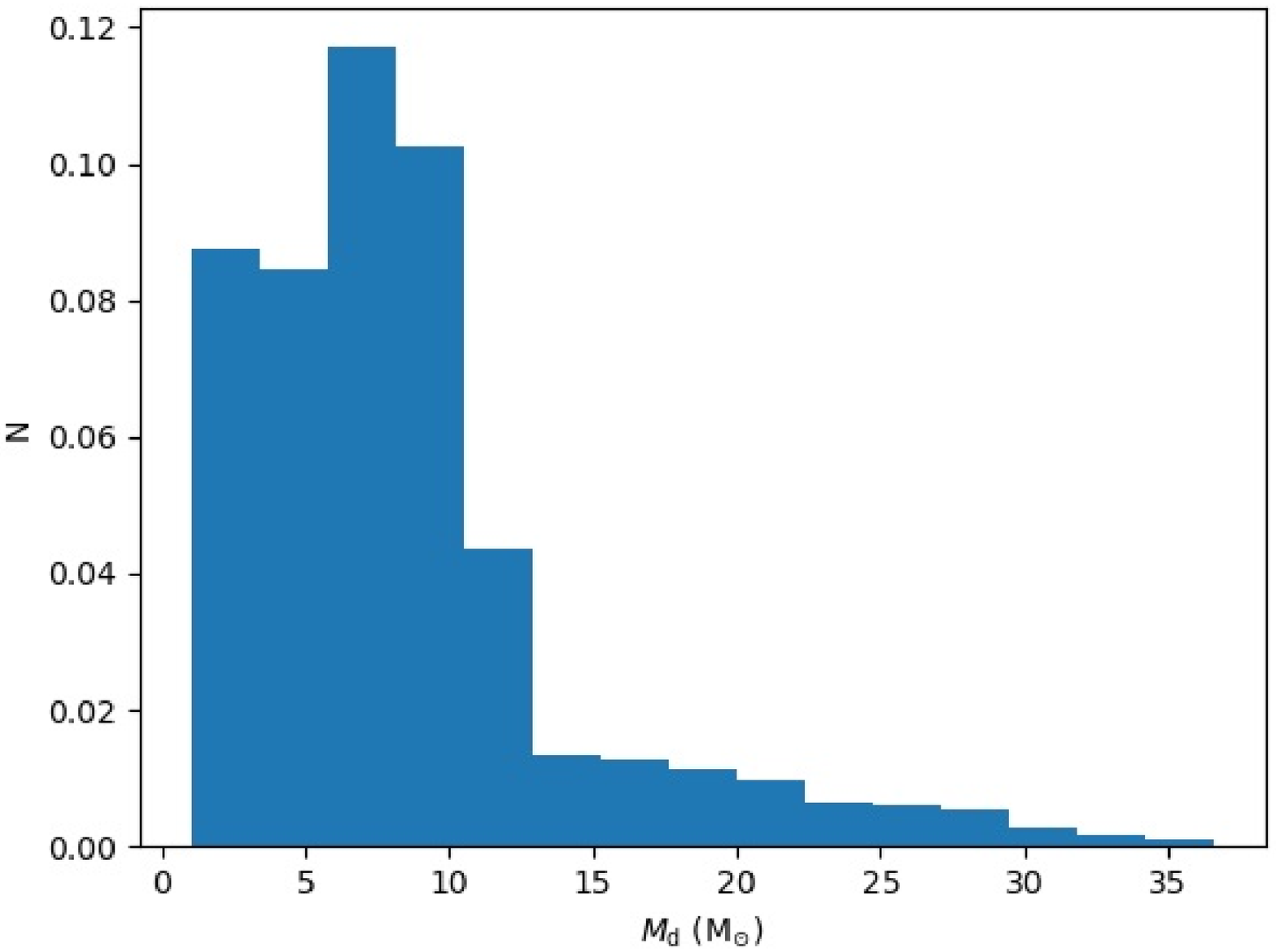}{0.4\textwidth}{}
          }
\caption{The distributions of the orbital parameters and the donor mass for NS-MS Roche-lobe overfilling XRBs with a mass transfer rate lower than $10^{-7}\ M_{\odot}\ \rm{yr}^{-1}$. The blue triangle shows the position of Cir X-1 in the $P_{\rm{orb}}-e$ plane. The top and bottom panels correspond to $\sigma_{\rm{k,CCS}}= 150$ and $300\ \rm{km}\ \rm{s}^{-1}$, respectively.\label{fig:3}}
\end{figure}

To guarantee that the mass transfer is not too high to lead to CE evolution, we set another criterion that the mass transfer rate should be lower than $10^{-7}\ M_{\odot}\ \rm{yr}^{-1}$. In the BSE code, the mass transfer is estimated in the following form \citep{T97}:
\begin{equation}
\dot{M}_{\rm{R}}=F(M)[{\rm ln}\ (R/R_{\rm{L}})]^{3}\ M_{\odot}\ \rm{yr^{-1}},
\end{equation}
where $F = 3\times10^{-6}[\rm{min}(M_{1},5.0)]^{2}$, $M$, $R$ and $R_{\rm{L}}$ are the mass, radius and Roche lobe radius of the donor, respectively. Figure \ref{fig:3} shows the distributions of the selected binaries that have a mass transfer rate lower than $10^{-7}\ M_{\odot}\ \rm{yr}^{-1}$ at periastron. The position of Cir X-1 disfavors the possibility that it has a main-sequence donor star. The birthrates with $\sigma_{\rm{k,CCS}}= 150$ and $300\ \rm{km}\ \rm{s}^{-1}$ are $9.83\times 10^{-6}$ and $5.05\times 10^{-6}\ \rm{yr}^{-1}$, respectively. With a rough estimation, we can probably observe one such XRB associated with an SNR in our galaxy.

\begin{figure}
\gridline{\fig{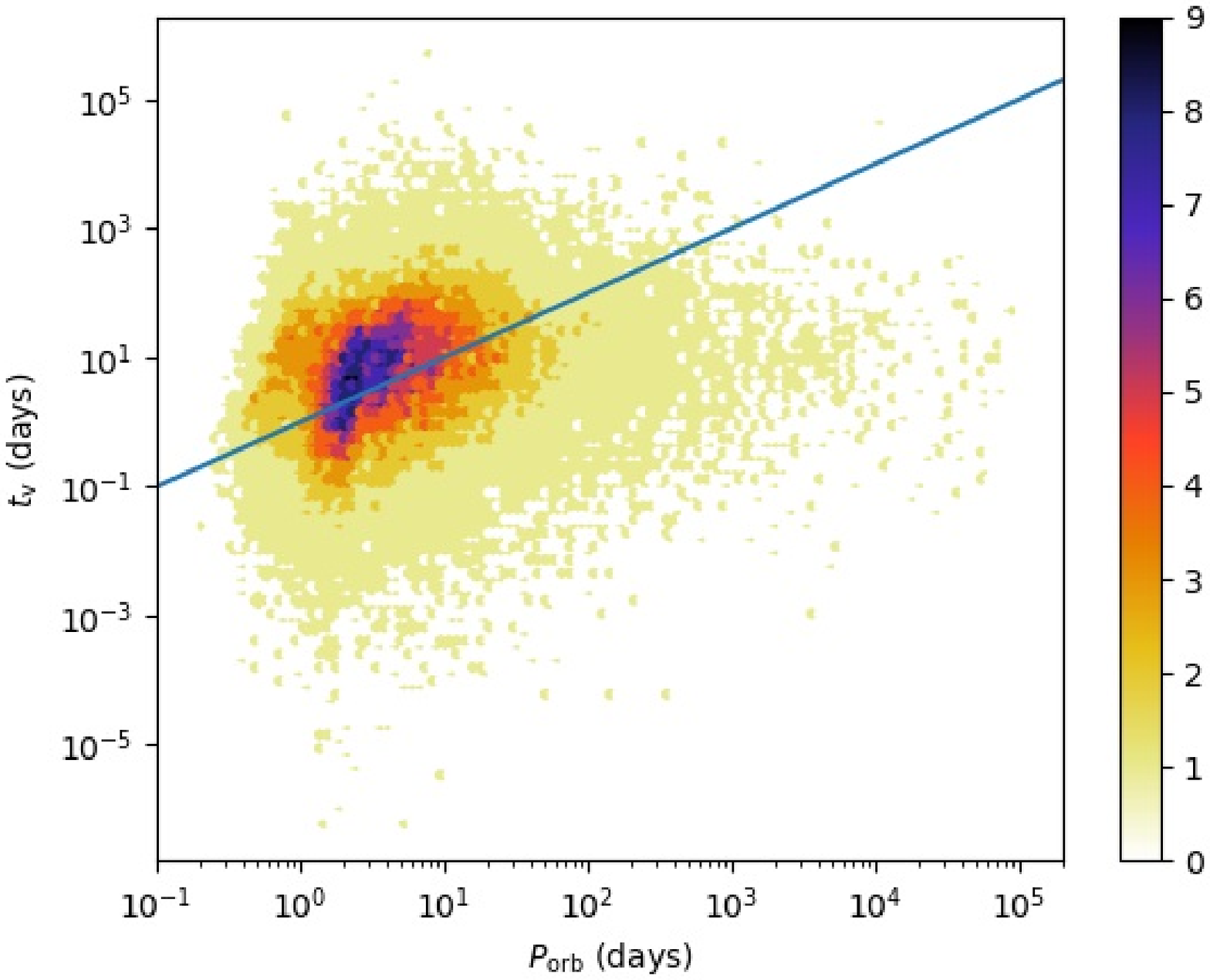}{0.4\textwidth}{}
          \fig{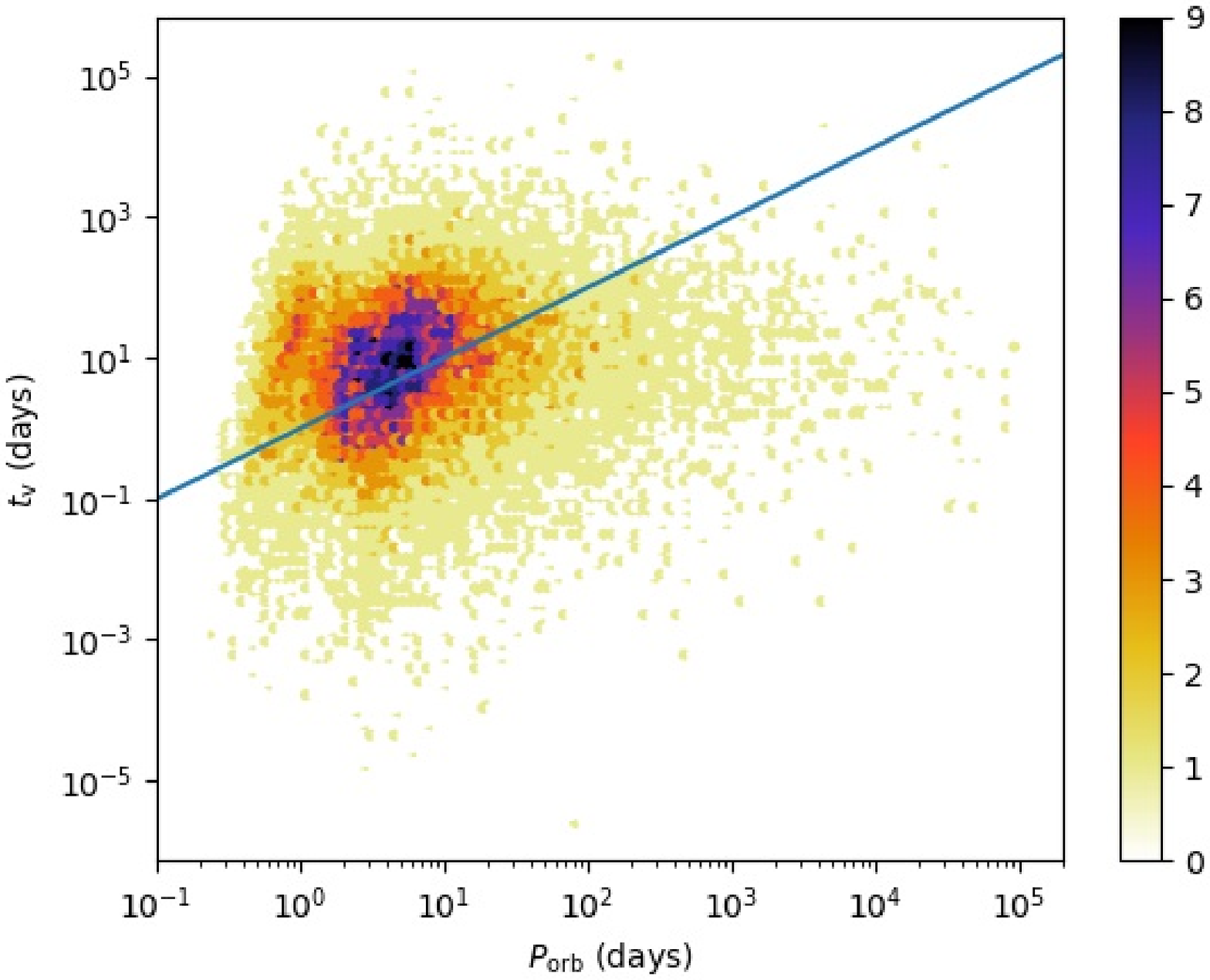}{0.4\textwidth}{}
          }
\gridline{\fig{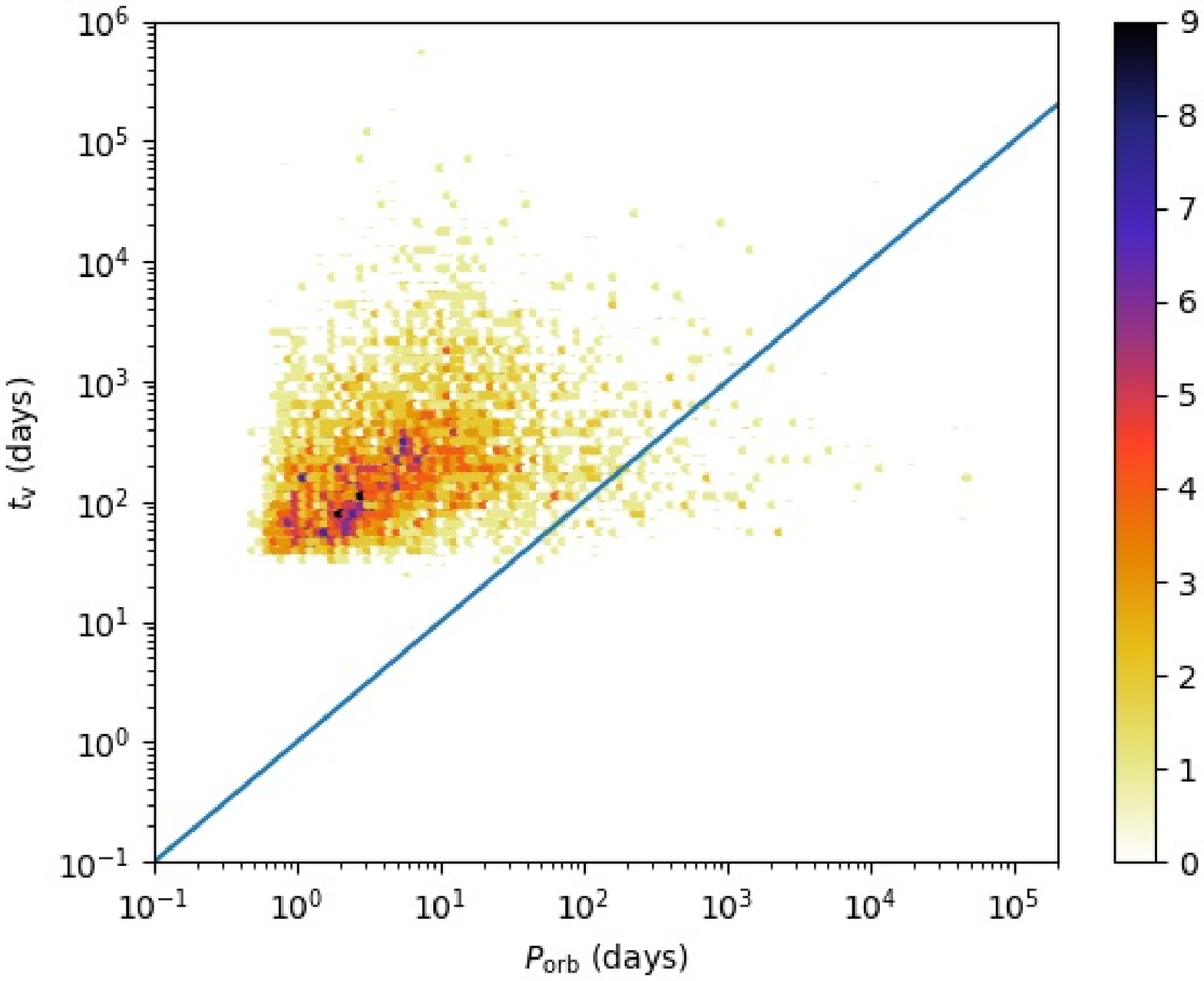}{0.4\textwidth}{}
          \fig{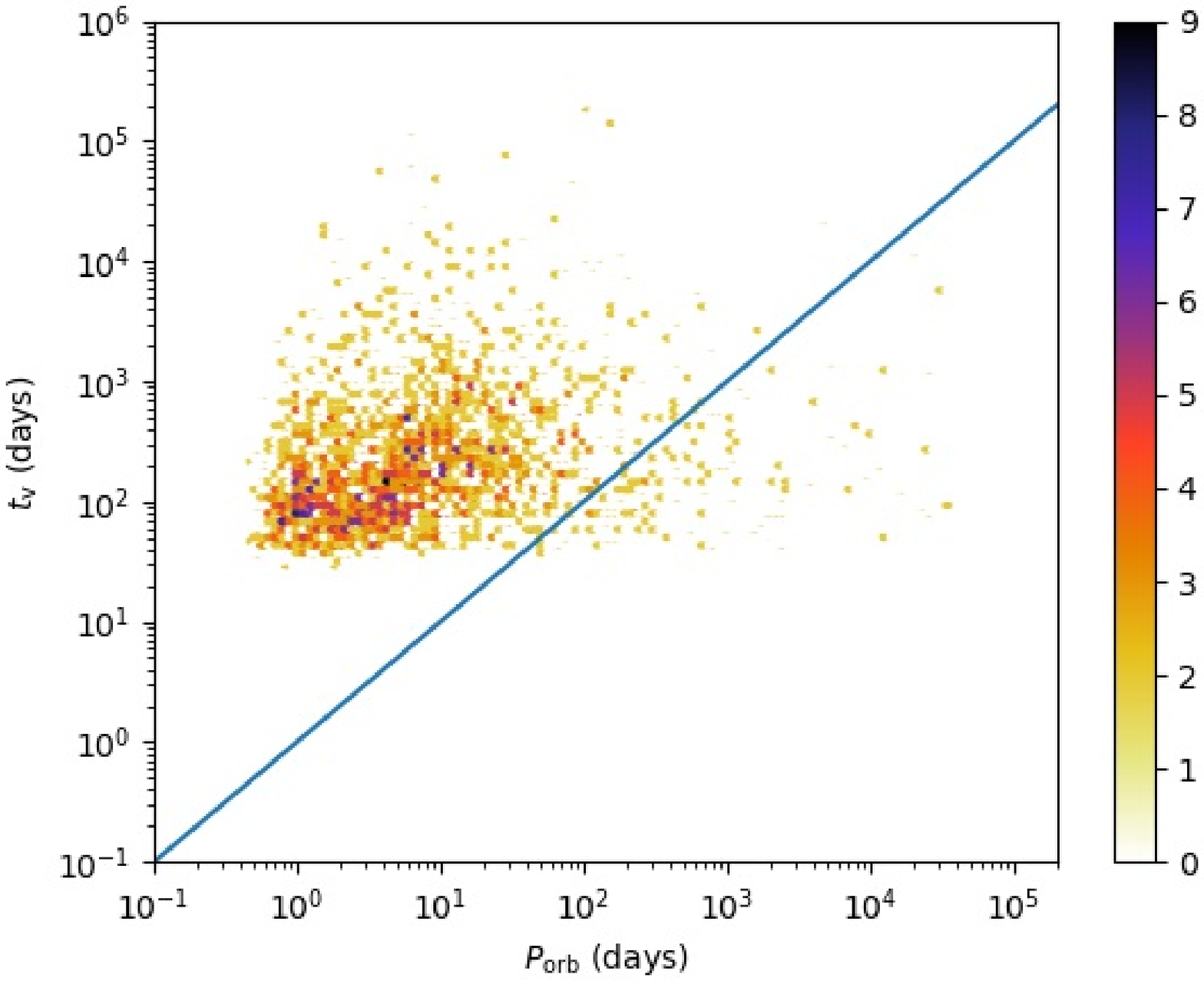}{0.4\textwidth}{}
          }
\caption{The distributions of NS-MS Roche-lobe overfilling XRBs within $10^{5}\ \rm{yr}$ after the SNe in the viscous timescale versus orbital period plane. The line represents the two values are equal. The top panels show all of the binaries and the bottom panels show the binaries with the mass transfer rate lower than $10^{-7}\ M_{\odot}\ \rm{yr}^{-1}$. The left and right panels correspond to $\sigma_{\rm{k,CCS}}= 150$ and $300\ \rm{km}\ \rm{s}^{-1}$, respectively.\label{fig:4}}
\end{figure}

Because the companion star overflows its Roche lobe only at periastron, the mass transfer may last for a short time, making the binaries be a transient source. So, we compare the viscous timescale of the accretion disk with the orbital period to see if the mass accretion is stable. If the viscous timescale is longer than the orbital period, we expect that mass accretion can proceed over the whole orbit, and the binary can appear as a persistent source. The viscous timescale depends on the mass transfer rate and the size of the accretion disk, and can be estimated to be \citep{F02}:
\begin{equation}
t_{\rm v} = R_{0}/v_{\rm{R}}\simeq 3\times10^{5}\alpha^{-4/5}(\dot M/10^{16}{\rm g\ s^{-1}})^{-3/10}m_{1}^{1/4}R_{0,10}^{5/4}\ \rm{s},
\end{equation}
where $\dot M$ is the mass transfer rate, $\alpha$ is the viscosity parameter, $m_{1}$ is the accretor mass in unit of solar mass, $R_{0}$ is the outer disk radius and $R_{0,10}$ is the outer disk radius in unit of $10^{10}\ \rm{cm}$, which is comparable to the Roche lobe radius of the NS. Figure \ref{fig:4} shows the distributions of the binaries in the viscous timescale versus orbital period plane. The top panels show all of the binaries and the bottom panels show the binaries with a mass transfer rate lower than $10^{-7}\ M_{\odot}\ \rm{yr}^{-1}$. Overall, more than half of the binaries have a longer viscous timescale than the orbital period. In most of the binaries with a relatively low mass transfer rate, the viscous timescale is longer than the orbital period, meaning that the mass accretion is stable in these systems.

\begin{figure}
\gridline{\fig{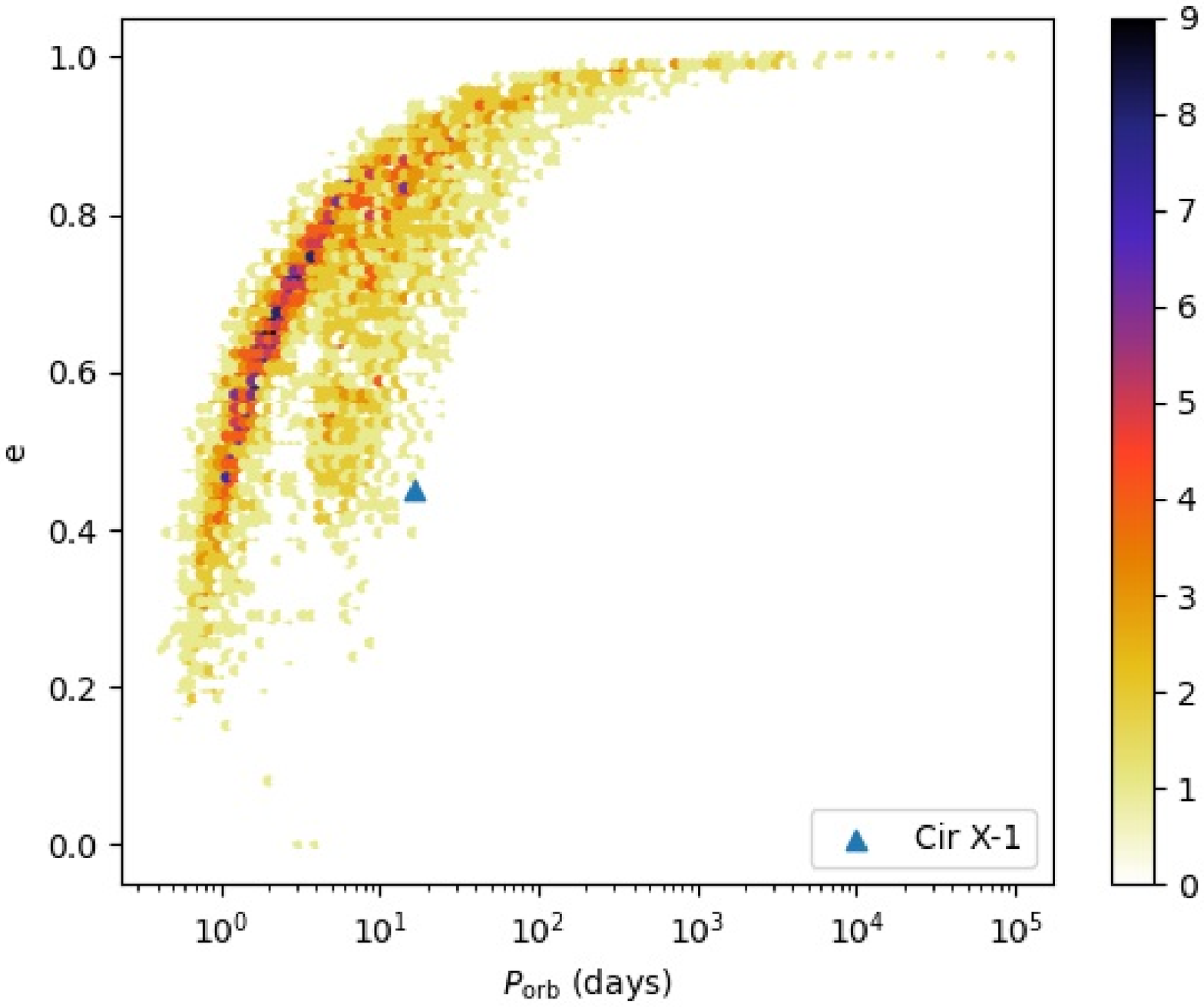}{0.3\textwidth}{}
          \fig{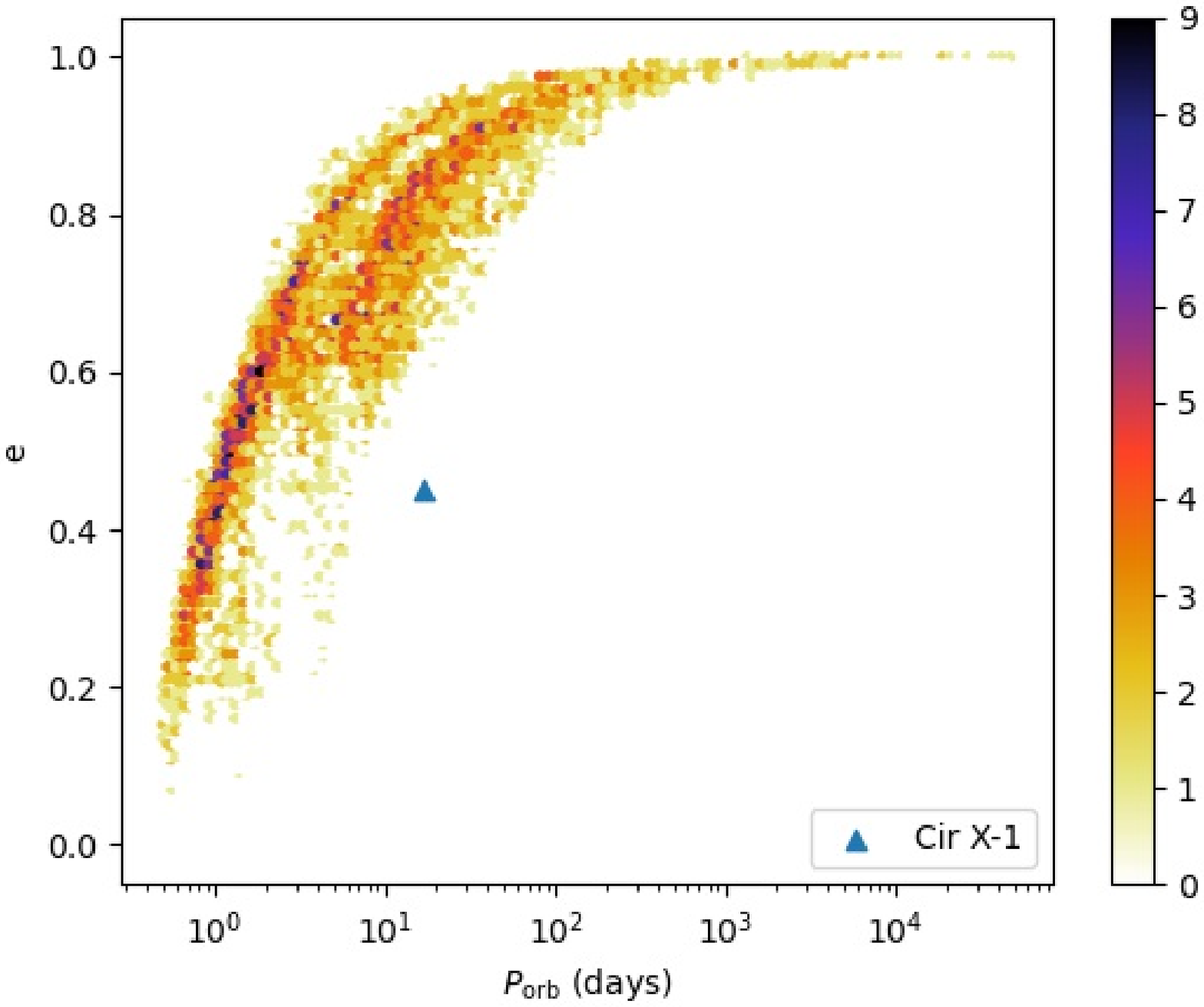}{0.3\textwidth}{}
          \fig{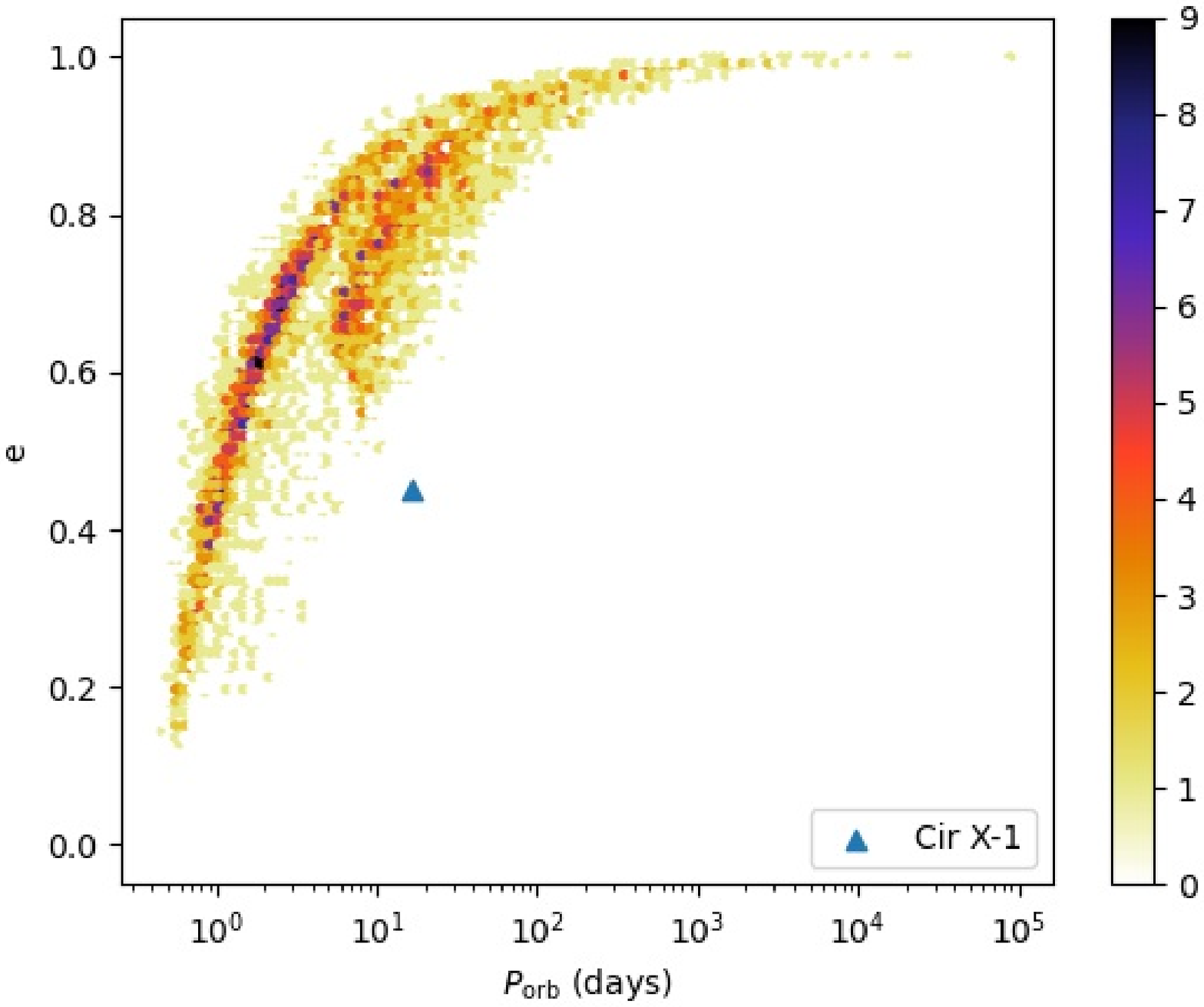}{0.3\textwidth}{}
          }
\caption{The $P_{\rm{orb}}-e$ distributions for NS-MS Roche-lobe overfilling XRBs with variations: $\alpha = 0.1$ (left), accretion efficiency of $0.5$ (middle), and the previous two combined (right). The blue triangle shows the position of Cir X-1.\label{fig:5}}
\end{figure}

Then, we consider variations in the common-envelope efficiency parameter $\alpha_{\rm{CE}}$, the mass accretion prescription, and the SFR in the galaxy. $\alpha_{\rm{CE}}$ represents the efficiency of transferring the orbital energy into the energy of the CE to unbind the envelope during CE evolution. We let $\alpha_{\rm{CE}} = 0.1 $, which makes it more difficult for successful envelope ejection, thus decreasing the number of binaries surviving the CE evolution. For the mass accretion prescription, we consider non-conservative mass accretion model with a constant accretion efficiency of $0.5$, which indicates that half of the transferred mass is accreted by the accretor. Comparing with the rotation-related accretion model, more mass would be accreted during the stable mass transfer phase and the critical mass ratios of the primary and the secondary stars for stable mass transfer decrease accordingly. Changes in the SFR will directly affect the birthrate of the binaries. Figure \ref{fig:5} shows the $P_{\rm{orb}}-e$ distributions of the binaries with different sets of parameters. A summary of the various parameters and the corresponding birthrates are presented in Table \ref{tab:2}.

\begin{deluxetable}{cccccc}
\tablecaption{Different Models and Corresponding Birthrates for NS-MS XRBs\label{tab:2}}


\tablehead{\colhead{Model number}& \colhead{$\sigma_{\rm{k,CCS}}$} & \colhead{$\alpha_{\rm{CE}}$} & \colhead{Accretion efficiency}  & \colhead{SFR} &\colhead{Birthrate} \\ 
\colhead{}& \colhead{$(\rm{km}\ \rm{s}^{-1})$} &\colhead{} & \colhead{}  & \colhead{$(M_{\odot}\ \rm{yr}^{-1})$}& \colhead{$(\rm{yr}^{-1})$}  } 

\startdata
1 & 150 & 1.0 & $(1-\frac{\omega}{\omega_{\rm{cr}}})$  & 3.0&$9.828\times10^{-6}$   \\
2 & 300 & 1.0 & $(1-\frac{\omega}{\omega_{\rm{cr}}})$  & 3.0&$5.052\times10^{-6}$  \\
3 & 150 & 0.1 & $(1-\frac{\omega}{\omega_{\rm{cr}}})$  & 3.0&$8.281\times10^{-6}$ \\
4 & 150 & 1.0 & 0.5  & 3.0&$8.893\times10^{-6}$   \\
5 & 150 & 0.1 & 0.5  & 3.0&$8.062\times10^{-6}$   \\
6 & 150 & 1.0 & $(1-\frac{\omega}{\omega_{\rm{cr}}})$  & 4.5&$1.474\times10^{-5}$   \\
\enddata

\end{deluxetable}

\subsection{Be X-ray Binaries}

\begin{figure}
\gridline{\fig{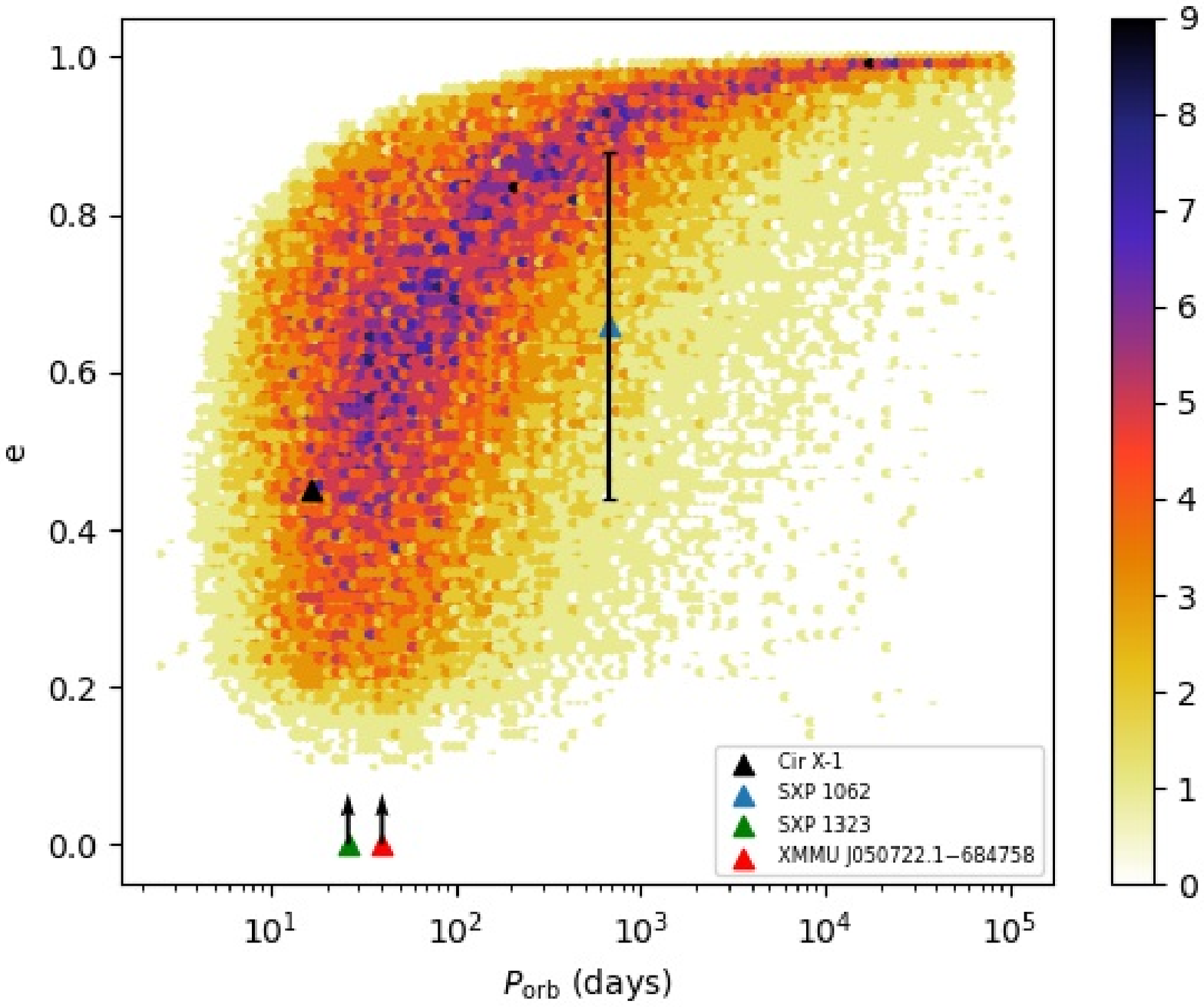}{0.3\textwidth}{}
          \fig{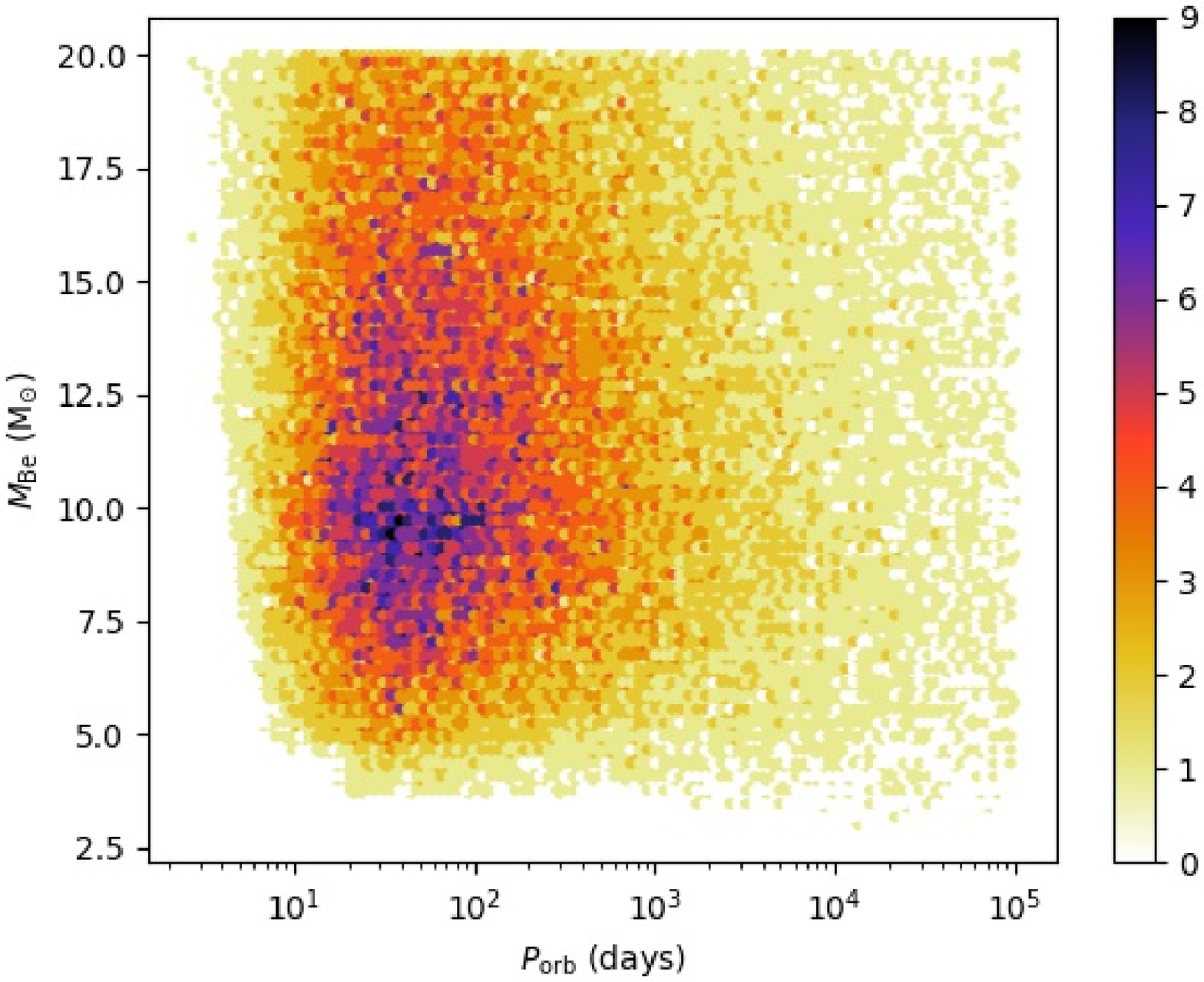}{0.3\textwidth}{}
          \fig{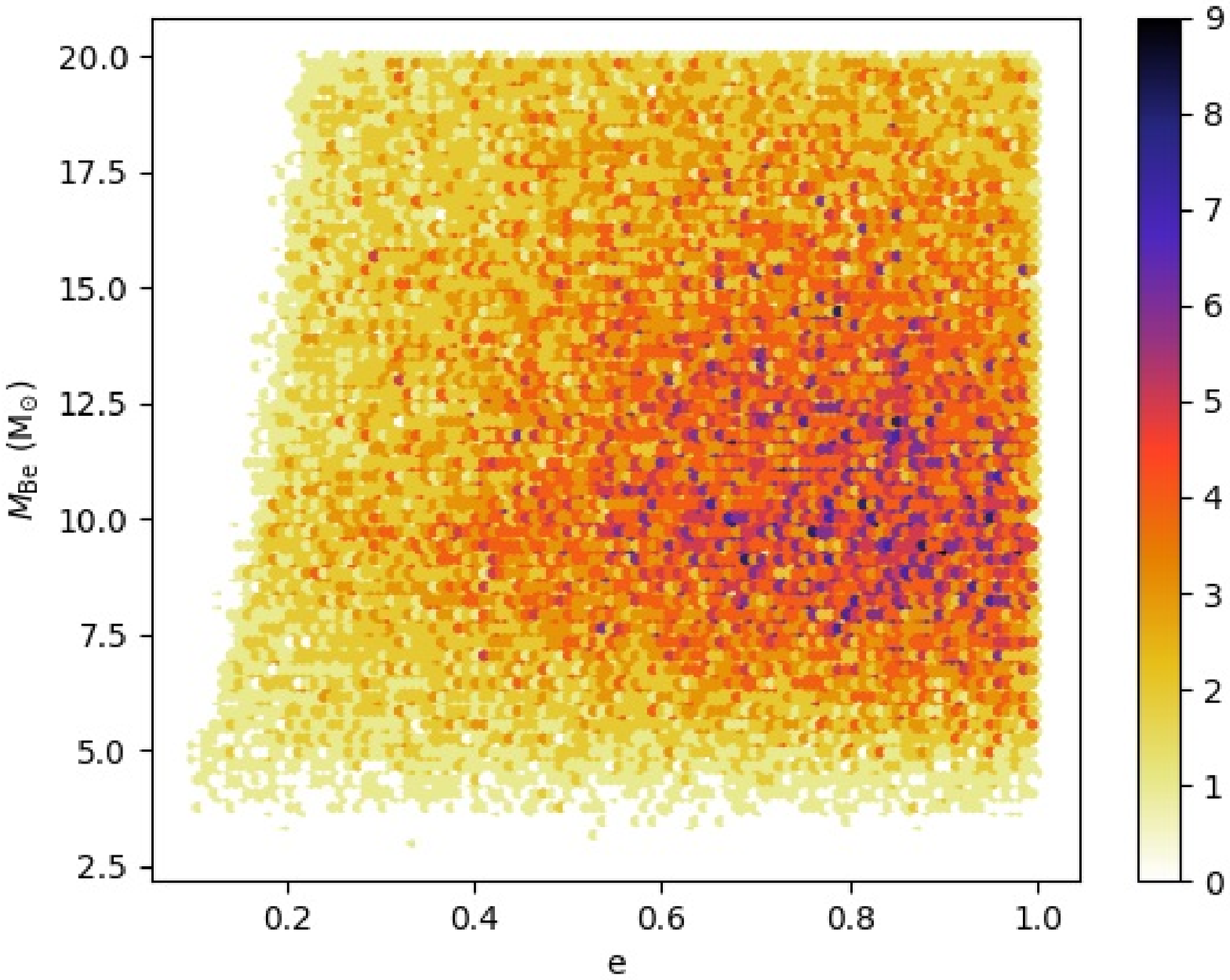}{0.3\textwidth}{}
          }
\gridline{\fig{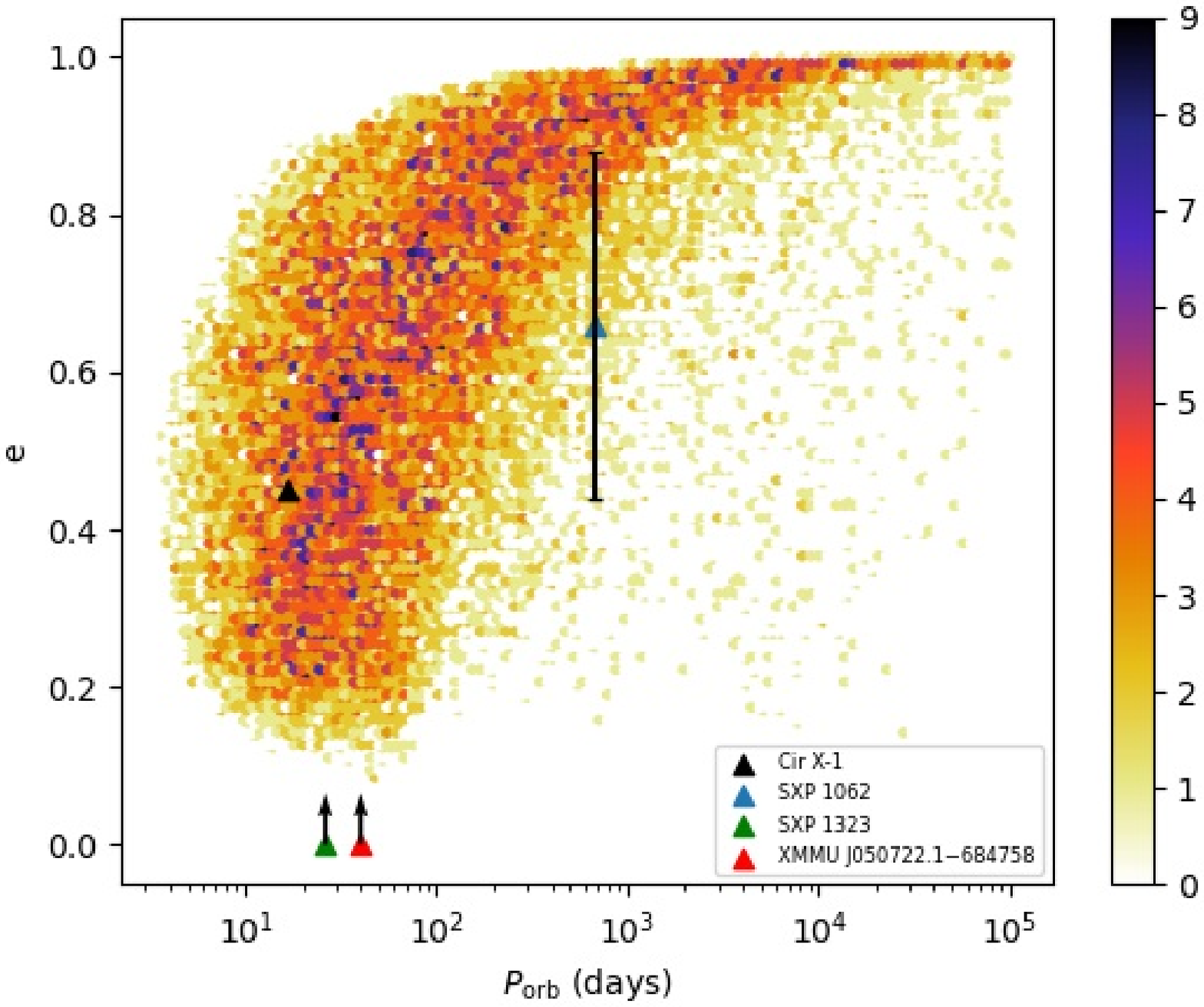}{0.3\textwidth}{}
          \fig{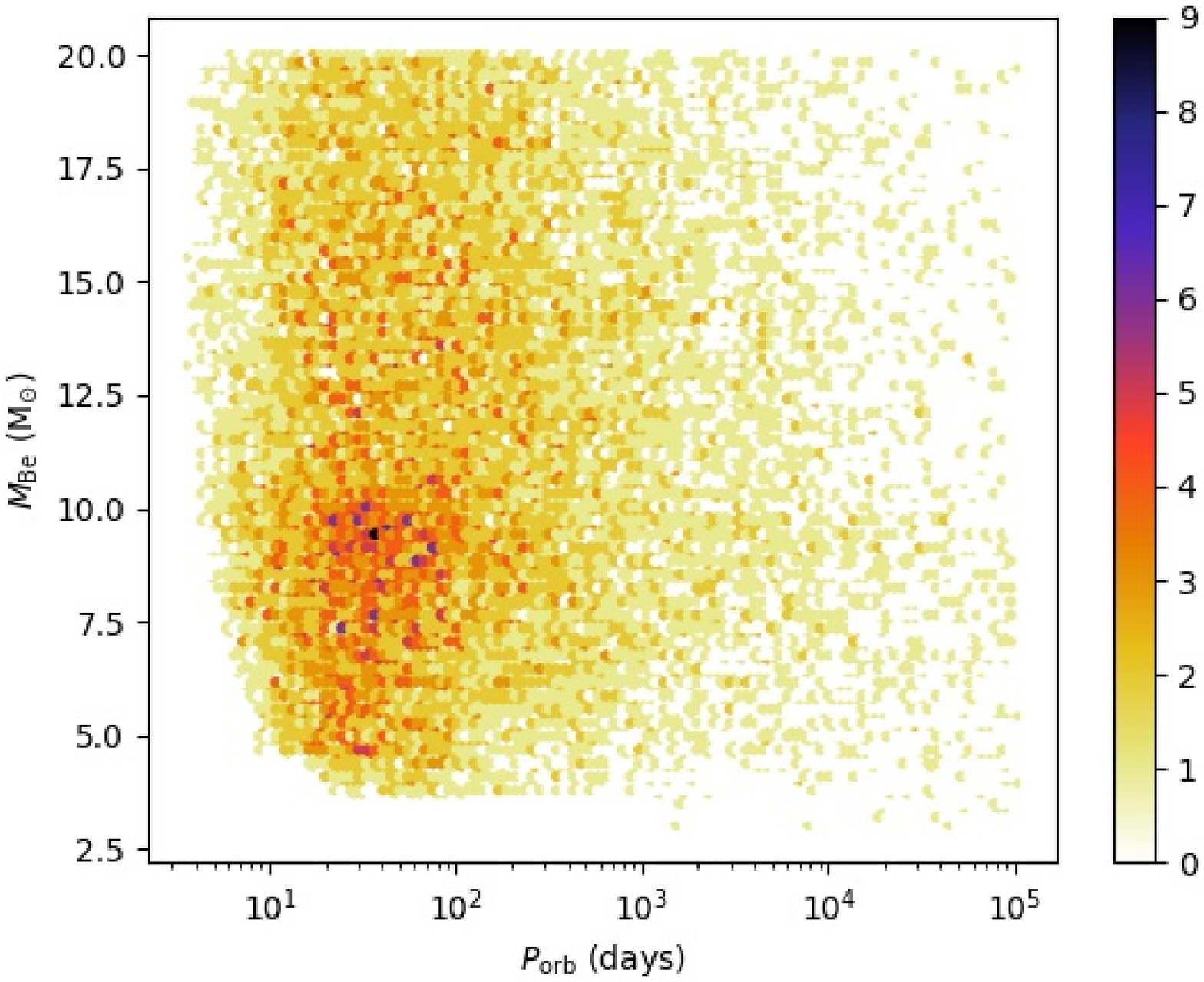}{0.3\textwidth}{}
          \fig{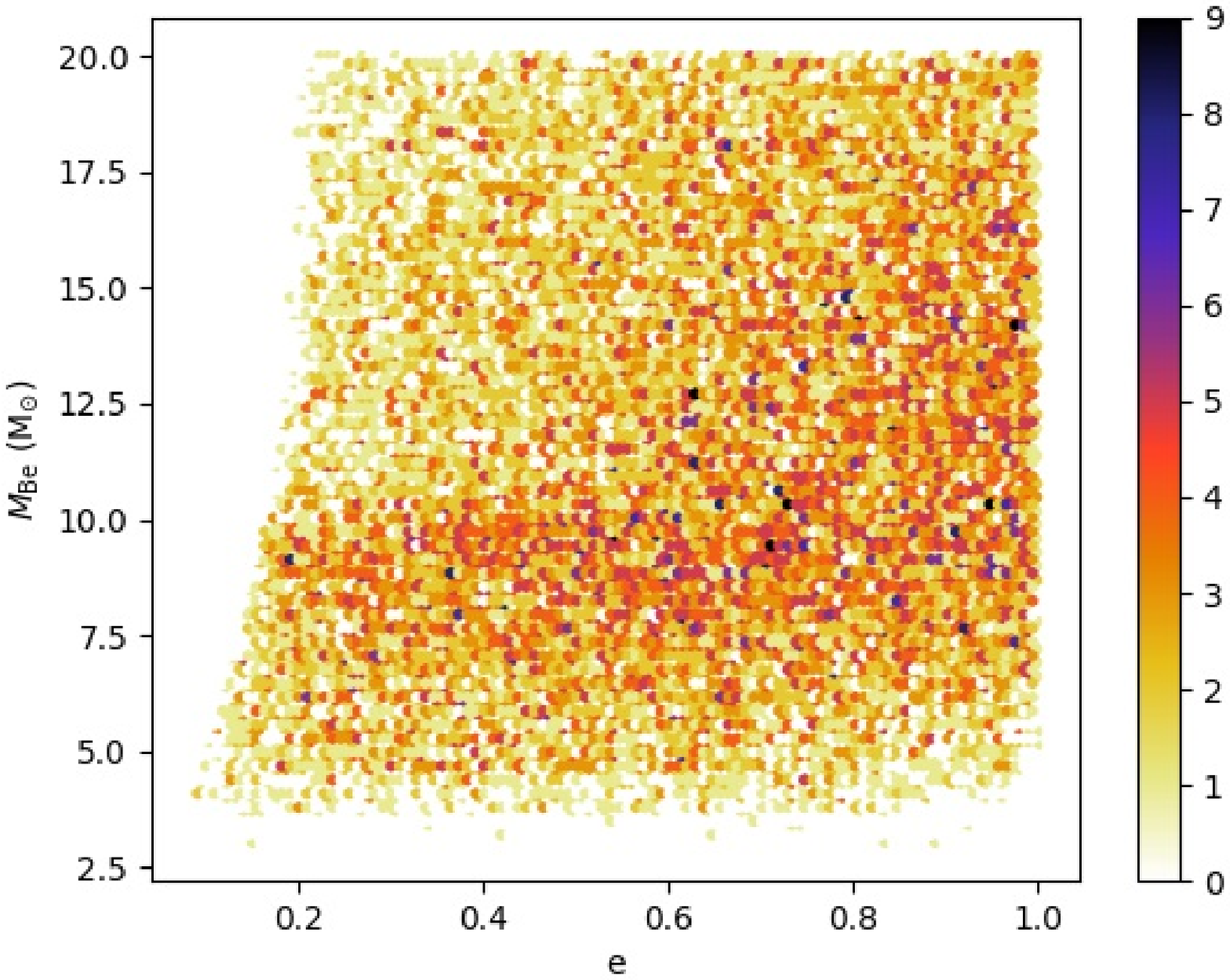}{0.3\textwidth}{}
          }
\caption{The distributions of parameters of BeXRBs. The left panel shows the $P_{\rm{orb}}-e$ plane, in which the black, blue, green, and red triangles indicate the positions of Cir X-1, SXP 1062, SXP 1323, and XMMU J050722.1-684758, respectively. The middle and right panels show their distribution in the $P_{\rm{orb}}-M_{\rm{Be}}$ plane and the $e-M_{\rm{Be}}$ plane, respectively. The top and bottom panels correspond to $\sigma_{\rm{k,CCS}}= 150$ and $300\ \rm{km}\ \rm{s}^{-1}$, respectively. \label{fig:6}}
\end{figure}

\begin{figure}
\gridline{\fig{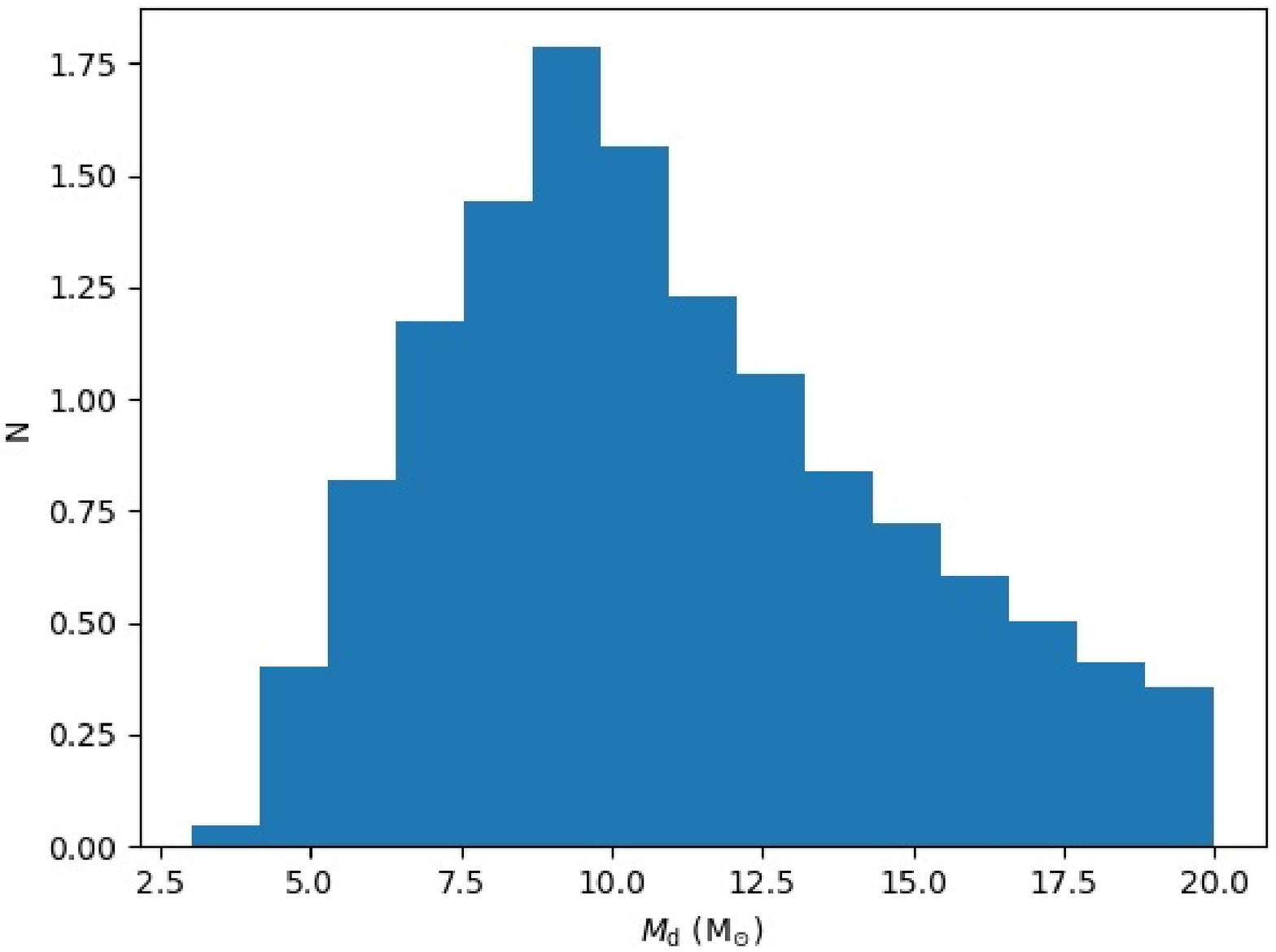}{0.4\textwidth}{}
          \fig{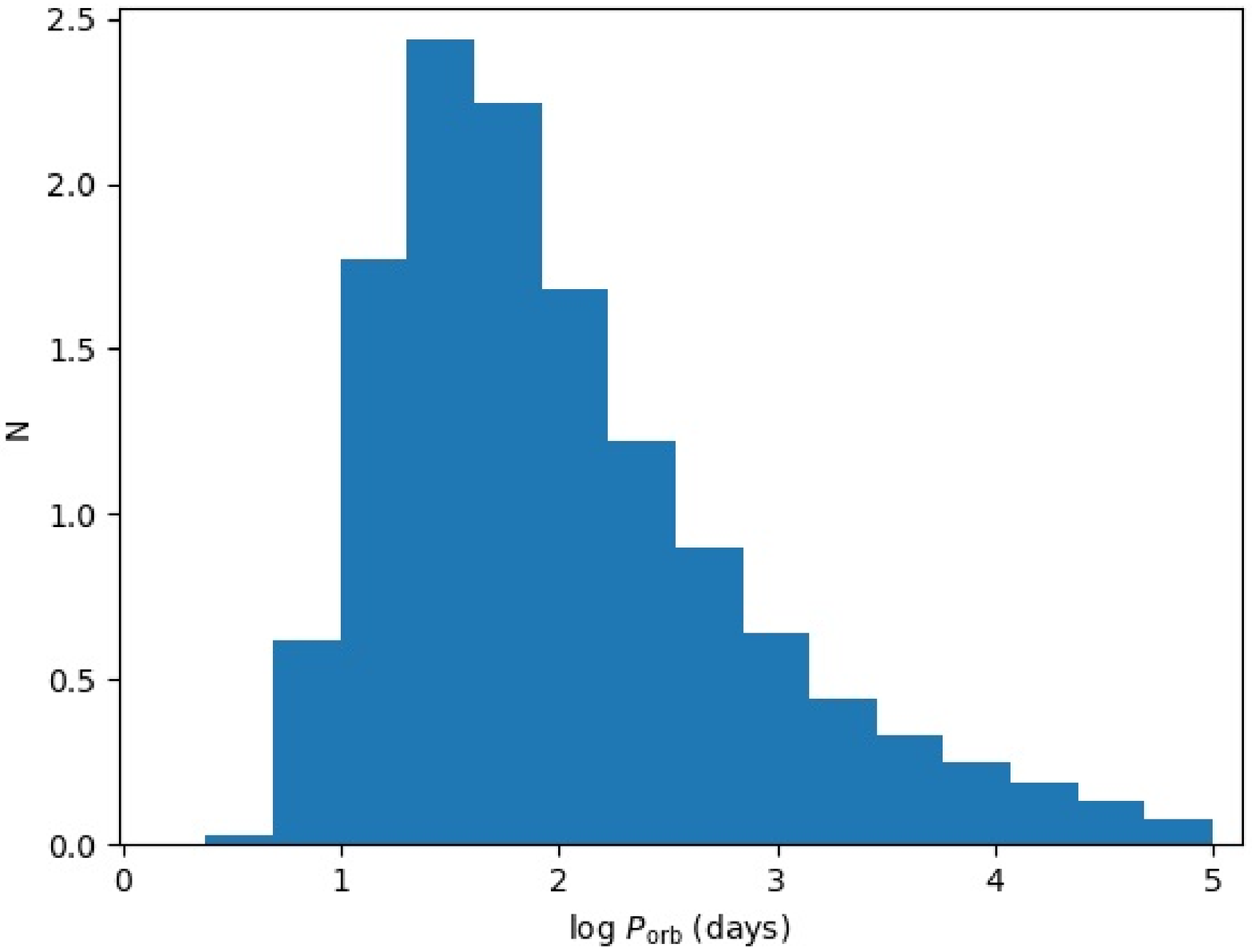}{0.4\textwidth}{}
          }
\gridline{\fig{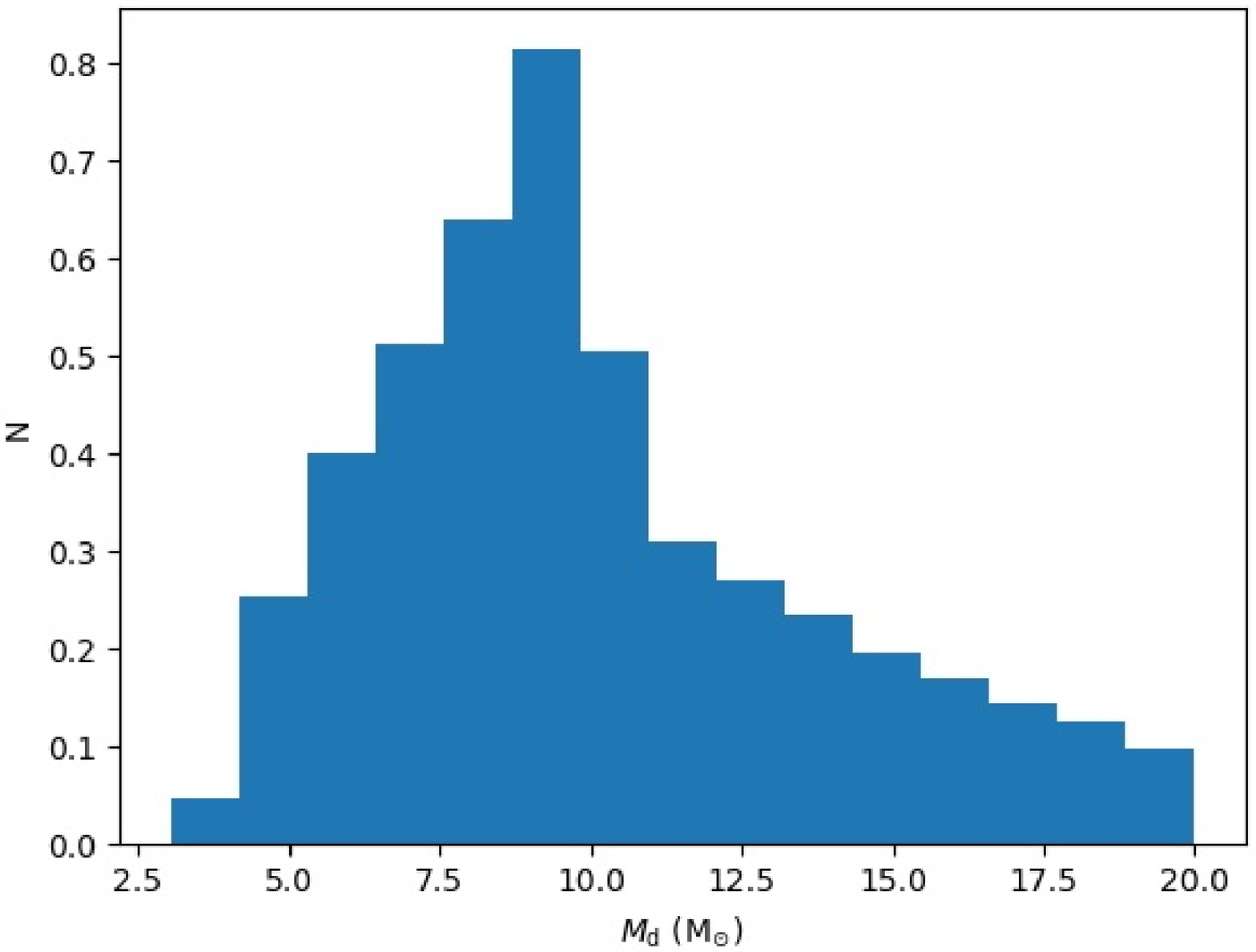}{0.4\textwidth}{}
          \fig{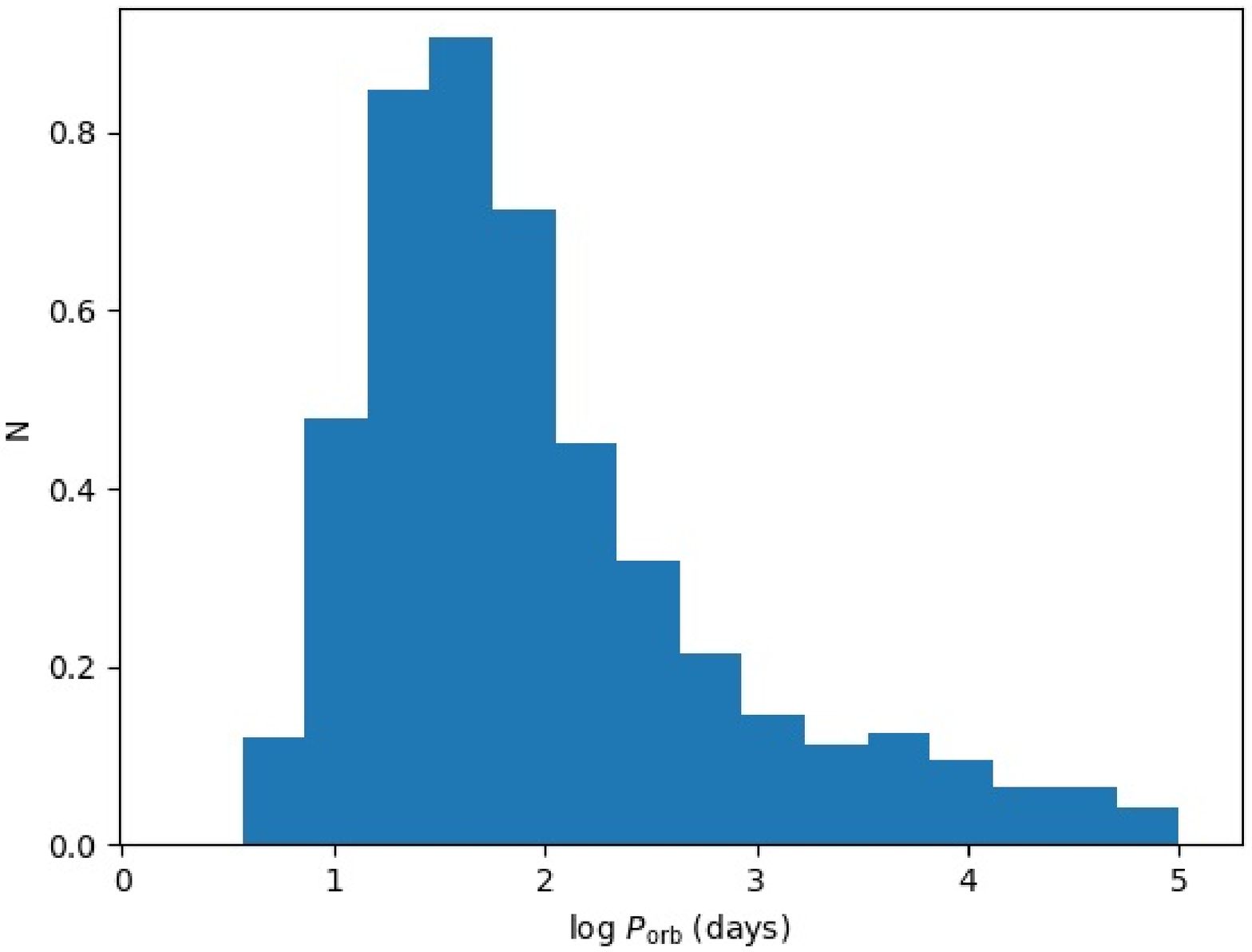}{0.4\textwidth}{}
          }
\caption{The distributions of the donor mass and the orbital period of BeXRBs. The top and bottom panels correspond to $\sigma_{\rm{k,CCS}}= 150$ and $300\ \rm{km}\ \rm{s}^{-1}$, respectively.\label{fig:7}}
\end{figure}

Now we consider the situation that the NS accretes from the circumstellar disk around a Be companion star. To define a Be star, we consider the companion star to be a main-sequence companion with mass between 3 and $20\ M_{\odot}$, as the Be phenomenon occur in stars with spectral type from early A to late O \citep{Ri13}. And we require that the companions are fast rotating stars with angular velocity higher than $80\%$ of the Keplerian limit at the equator \citep{Ri13}, which allow the formation of decretion disks. In addition, the companion stars are not overflowing their Roche lobes at periastron to avoid the destruction of the disks \citep{Pa10}. For the formation channel of BeXRBs, we adopt the result of \citet{Y14} that most of the primordial binaries have undergone stable mass transfer without CE evolution. Recently, \citet{V20} obtained a similar result that systems evolved through CE evolution are disfavored to produce BeXRBs because they do not match the observed distribution of the companion mass in BeXRBs. Then, we compare the size of the decretion disk with the Roche lobe radius of the companion. To estimate the size of the circumstellar disk, we use the truncation radius described in \citet{Zh04}:
\begin{equation}
R_{\rm{trunc}} = n^{-2/3}[GM_{\rm{Be}}/(2\pi/P_{\rm{orb}})^{2}]^{1/3},
\end{equation}
where $M_{\rm{Be}}$ is the mass of the Be star, $P_{\rm{orb}}$ is the orbital period, and we set $n = 3$ in the calculation. The idea of the truncation radius is based on the theoretical works on disk truncation in BeXRBs \citep{O01,O02}. The NS exerts a tidal torque on the decretion disk surrounding the Be star, causing the disk to be truncated at a certain radius, since the viscous torque in the disk becomes smaller than the tidal torque. As a result, the truncation radius can be approximately regarded as the size of the decretion disk. If $R_{\rm{trunc}}$ is larger than the Roche lobe radius of the Be star, the NS can accrete material from the truncated disk and the binary may appear as a BeXRB.
Figure \ref{fig:6} shows the distribution of the selected BeXRBs, from left to right, in the $P_{\rm{orb}}-e$, $P_{\rm{orb}}-M_{\rm{Be}}$, and $e-M_{\rm{Be}}$ planes and the top and bottom panels correspond to $\sigma_{\rm{k,CCS}}= 150$ and $300\ \rm{km}\ \rm{s}^{-1}$, respectively. In the $P_{\rm{orb}}-e$ panel, we can see that the binaries are mainly distributed in the region with the orbital periods within $10-200\ \rm{d}$ and the eccentricities larger than $0.2$. For binaries with orbital periods of few days, the companion stars are likely to fill the Roche lobe and hinder the formation of the decretion disks. The black, blue, green, and red triangles show the positions of Cir X-1, SXP 1306, SXP 1323 and XMMU J050722.1-684758 in the $P_{\rm{orb}}-e$ panel, respectively. If Cir X-1 is a BeXRB, it would be natural to explain its current orbital parameters. The $P_{\rm{orb}}-M_{\rm{Be}}$ and $e-M_{\rm{Be}}$ panels show that the companion masses are distributed around $10\ M_{\odot}$ while the eccentricity has a rather flat distribution. The overall distributions of the donor mass and the orbital period are shown in Figure \ref{fig:7}, which are compatible with the observation of current BeXRBs in the Milky Way and Magellanic Clouds. 

These young BeXRBs can exist and sustain mass transfer for a time as long as their main-sequence lifetimes. In our calculations, the birthrates of these systems with $\sigma_{\rm{k,CCS}}= 150$ and $300\ \rm{km}\ \rm{s}^{-1}$ are $1.30\times 10^{-4}$ and $4.70\times 10^{-5}\ \rm{yr}^{-1}$, respectively. Assuming that SNRs can last for about $10^{5}\ \rm{yr}$, we expect that there could be a few BeXRBs that are harbored in SNRs in the Milky Way.

\begin{figure}
\gridline{\fig{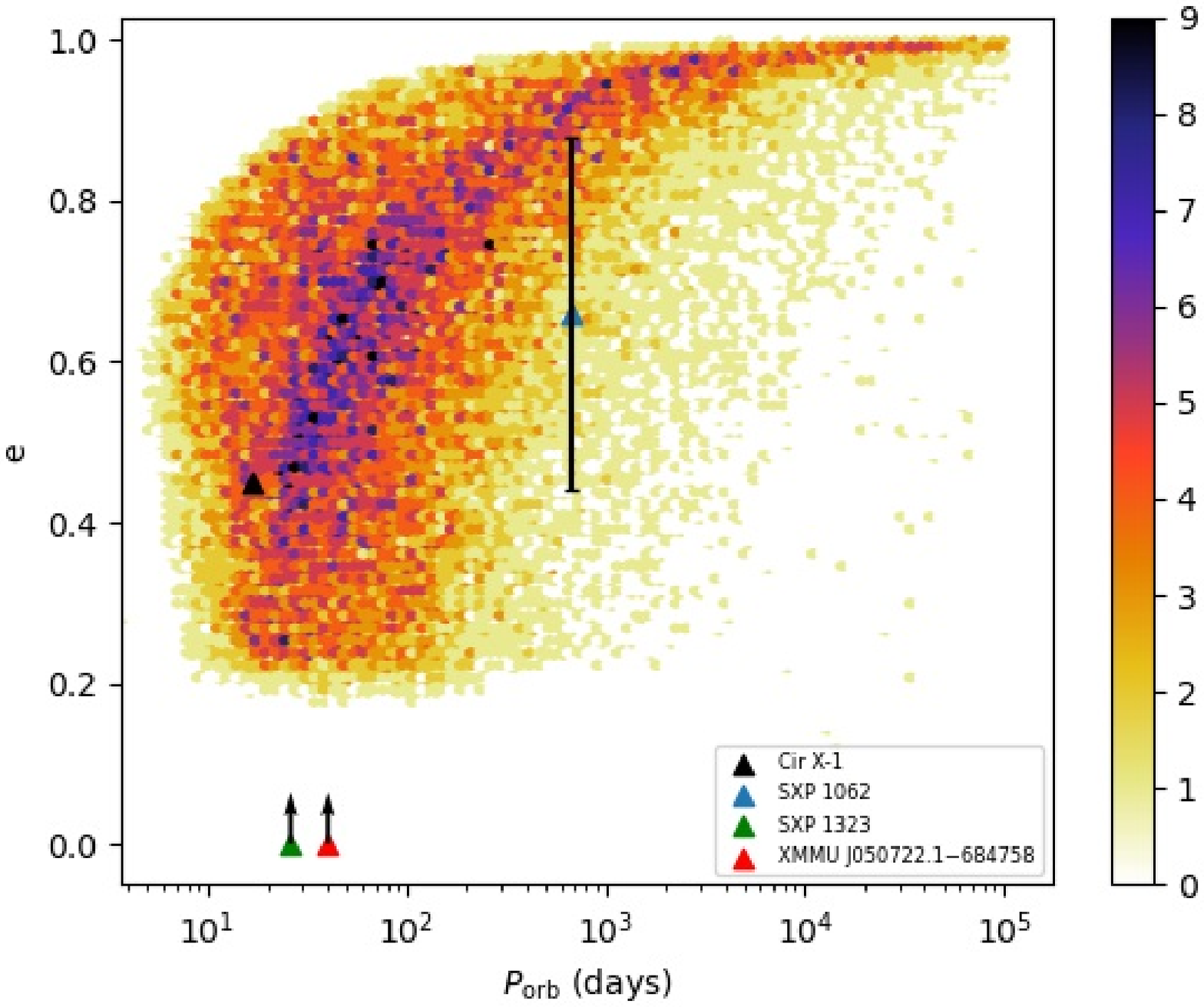}{0.3\textwidth}{}
          \fig{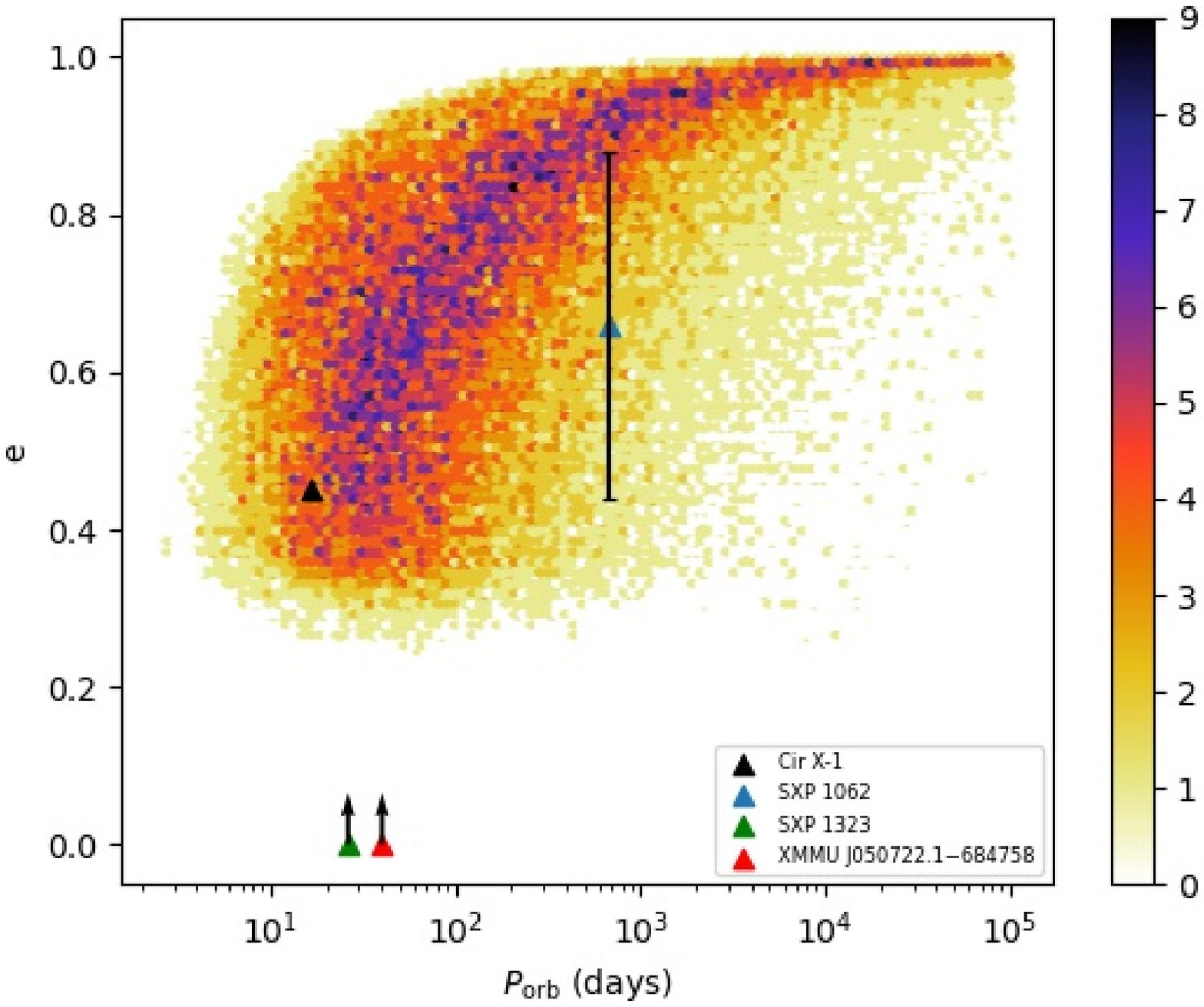}{0.3\textwidth}{}
          \fig{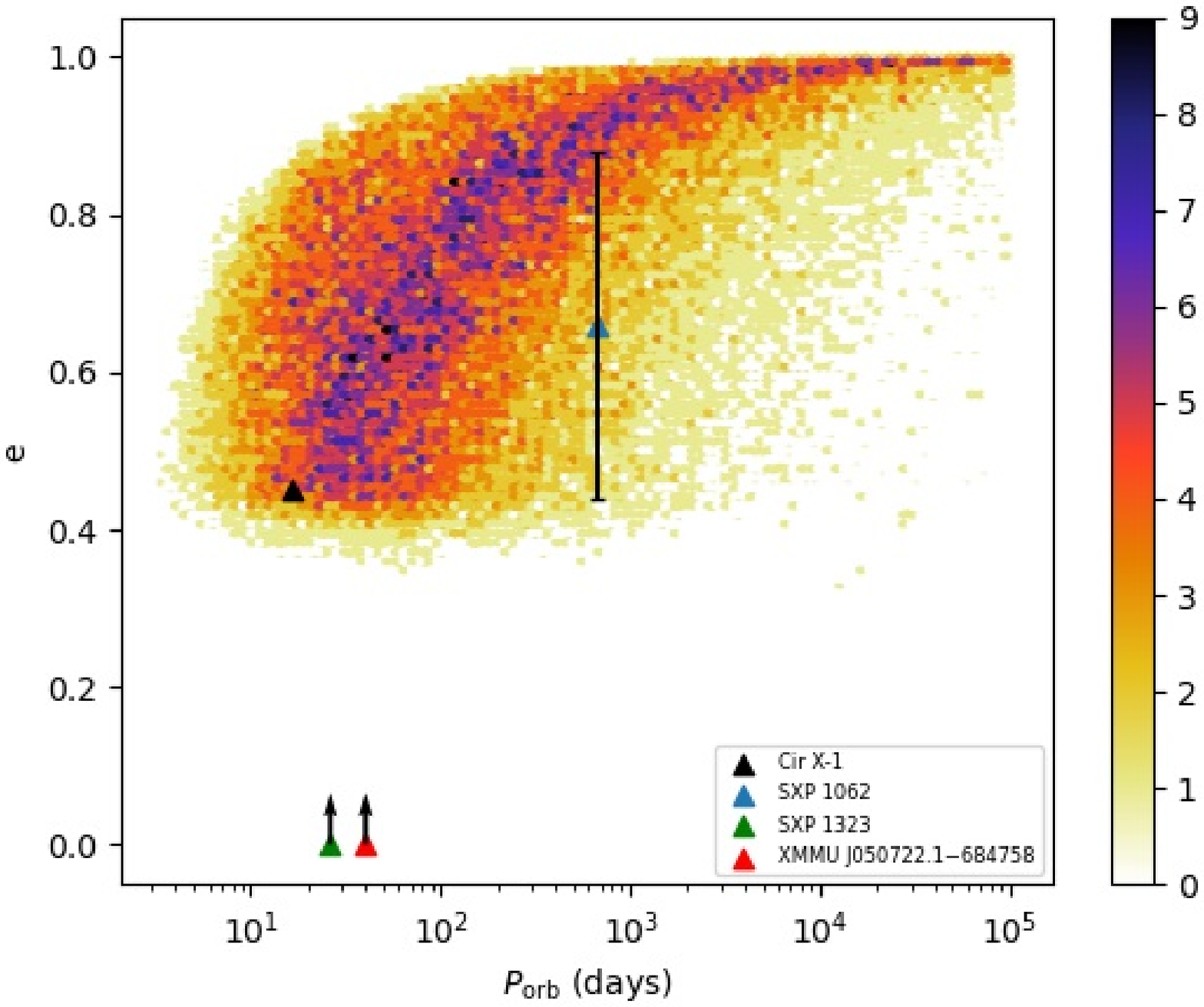}{0.3\textwidth}{}
          }
\caption{The $P_{\rm{orb}}-e$ distributions of BeXRBs with variations: accretion efficiency of $0.5$ (left), truncation parameter $n = 4$ (middle), and $n = 5$ (right). The black, blue, green, and red triangles indicate the positions of Cir X-1, SXP 1062, SXP 1323, and XMMU J050722.1-684758, respectively.\label{fig:8}}
\end{figure}

\begin{figure}
\plottwo{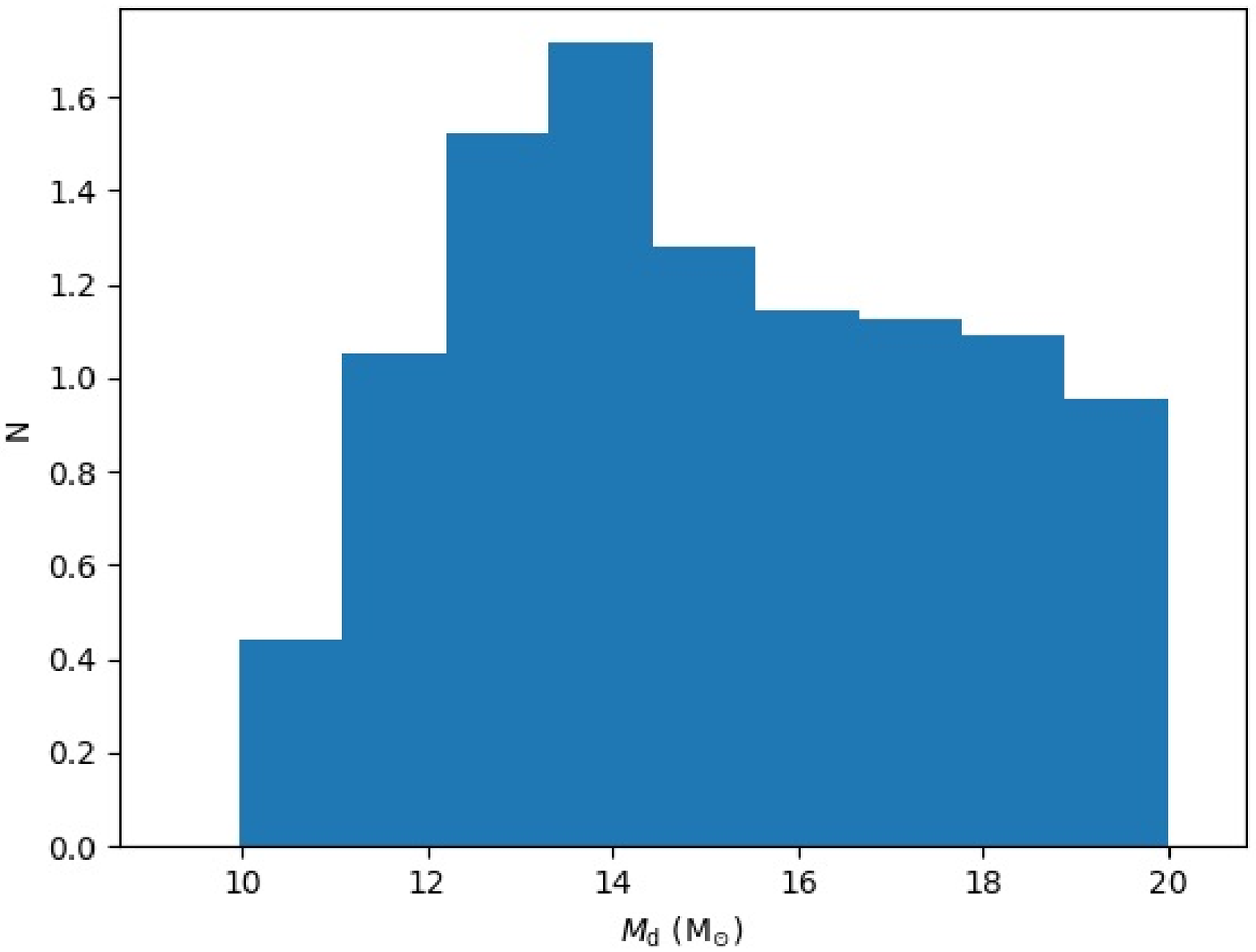}{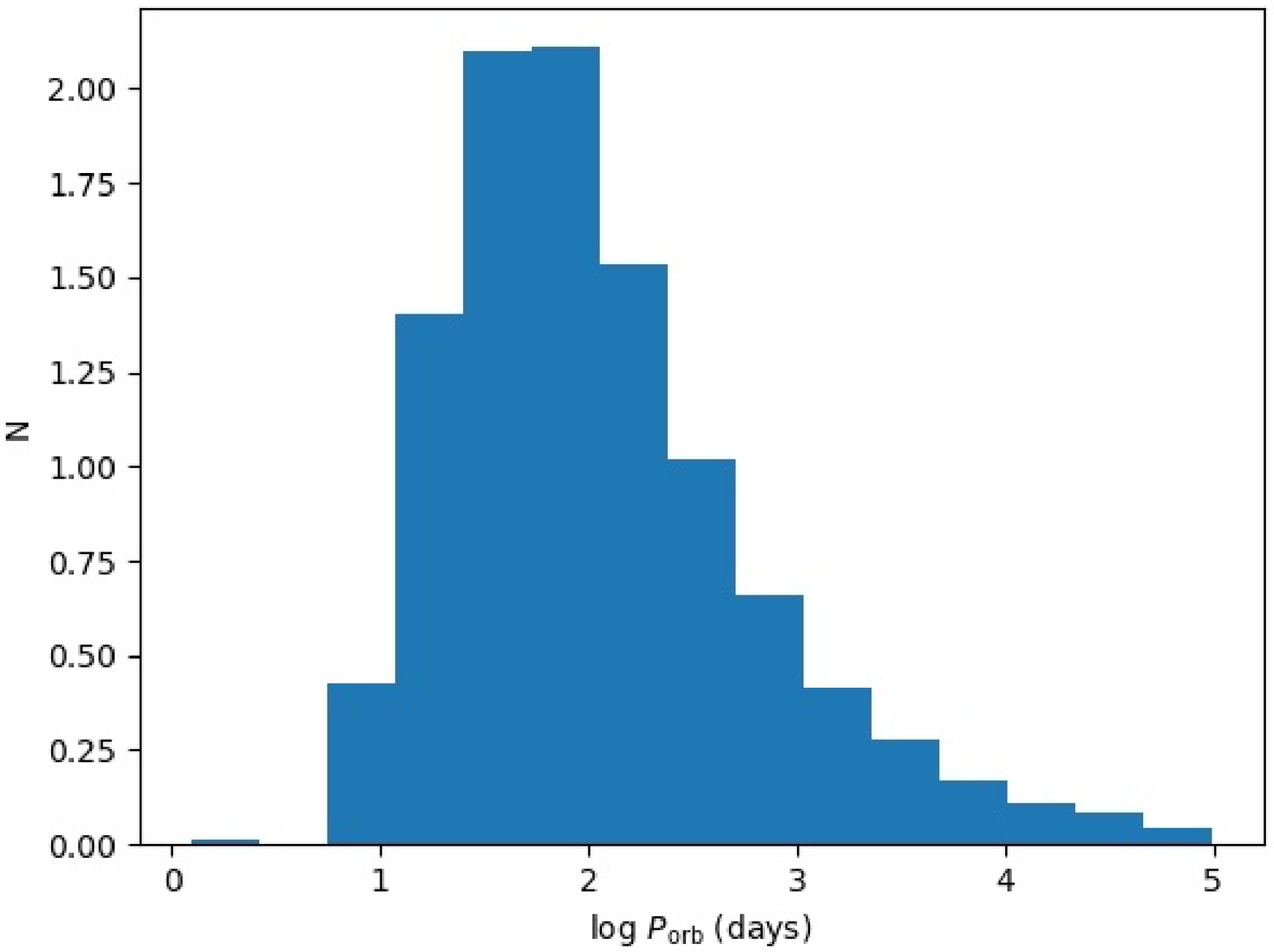}
\caption{The distributions of the donor mass and the orbital period of BeXRBs under the accretion model with an efficiency of $0.5$.\label{fig:9}}
\end{figure}

We then consider the influence of the variations in the truncation parameter $n$, the accretion efficiency, and the SFR. The parameter $n$, which is related to the Shakura-Sunyaev viscosity parameter $\alpha$, indicates the resonance radius and determines the size of the decretion disk. According to \citet{O02}, $n=3$ if $0.17\lesssim\alpha\lesssim0.7$ and $n=4$ if $0.037\lesssim\alpha\lesssim0.17$. Figure \ref{fig:8} shows the $P_{\rm{orb}}-e$ distributions of the BeXRBs with different values of parameters, and Figure \ref{fig:9} shows the distributions of the Be star mass and the orbital period for the accretion efficiency of $0.5$. We can see with larger $n$, namely smaller decretion disk size, it is harder for the binaries with low eccentricities to meet the requirement. And if we adopt the accretion efficiency of $0.5$ rather than rotation-dependent, the Be star would become more massive by accreting more mass from the hydrogen-rich primary through stable mass transfer. Table \ref{3} summarizes the various parameters adopted and the corresponding birthrates for BeXRBs.

From the perspective of observation, we do not know whether the Be phenomenon is universal for all of the main-sequence companions, so the estimated numbers in our calculations can be the upper limit for BeXRBs. If we consider that the SNRs exist for $10^{4}\ \rm{yr}-10^{5}\ \rm{yr}$ and adopt different birthrates calculated with the variations, we can estimate that in the Milky Way, the number of BeXRBs could be $0.47-19$.

\begin{deluxetable}{cccccc}
\tablecaption{Different Models and Corresponding Birthrates for BeXRBs\label{3}}


\tablehead{\colhead{Model number}& \colhead{$\sigma_{\rm{k,CCS}}$} & \colhead{n} & \colhead{Accretion efficiency}  & \colhead{SFR} &\colhead{Birthrate} \\ 
\colhead{}& \colhead{$(\rm{km}\ \rm{s}^{-1})$} &\colhead{} & \colhead{}  & \colhead{$(M_{\odot}\ \rm{yr}^{-1})$}& \colhead{$(\rm{yr}^{-1})$}  } 

\startdata
1 & 150 & 3  & $(1-\frac{\omega}{\omega_{\rm{cr}}})$& 3.0 &$1.295\times10^{-4}$   \\
2 & 300 & 3  & $(1-\frac{\omega}{\omega_{\rm{cr}}})$& 3.0 &$4.702\times10^{-5}$  \\
3 & 150 & 4  & $(1-\frac{\omega}{\omega_{\rm{cr}}})$& 3.0 &$1.148\times10^{-4}$ \\
4 & 150 & 5  & $(1-\frac{\omega}{\omega_{\rm{cr}}})$& 3.0 &$1.022\times10^{-4}$   \\
5 & 150 & 3  & 0.5 & 3.0& $1.059\times10^{-4}$   \\
6 & 150 & 3  & $(1-\frac{\omega}{\omega_{\rm{cr}}})$& 4.5 & $1.943\times10^{-4}$   \\
\enddata

\end{deluxetable}
\subsection{Supergiant X-ray binaries}

Finally, we consider the case that the companion stars are supergiant stars. They have much shorter lifetime compared with main-sequence stars of the same mass, thus the number of SGXBs within SNRs in the calculations is supposed to be much smaller than that of XRBs with main-sequence companions, and so is the birthrate. Although most of known SGXBs are wind-fed systems, there are several SGXBs in which mass transfer proceeds via Roche-lobe overflow. These disk-fed systems will go through rapid orbital evolution due to the extreme mass ratio and their lifespan may be very short. In this case, they may have an X-ray lifetime much shorter than the lasting time of the SNRs.

\begin{figure}
\plottwo{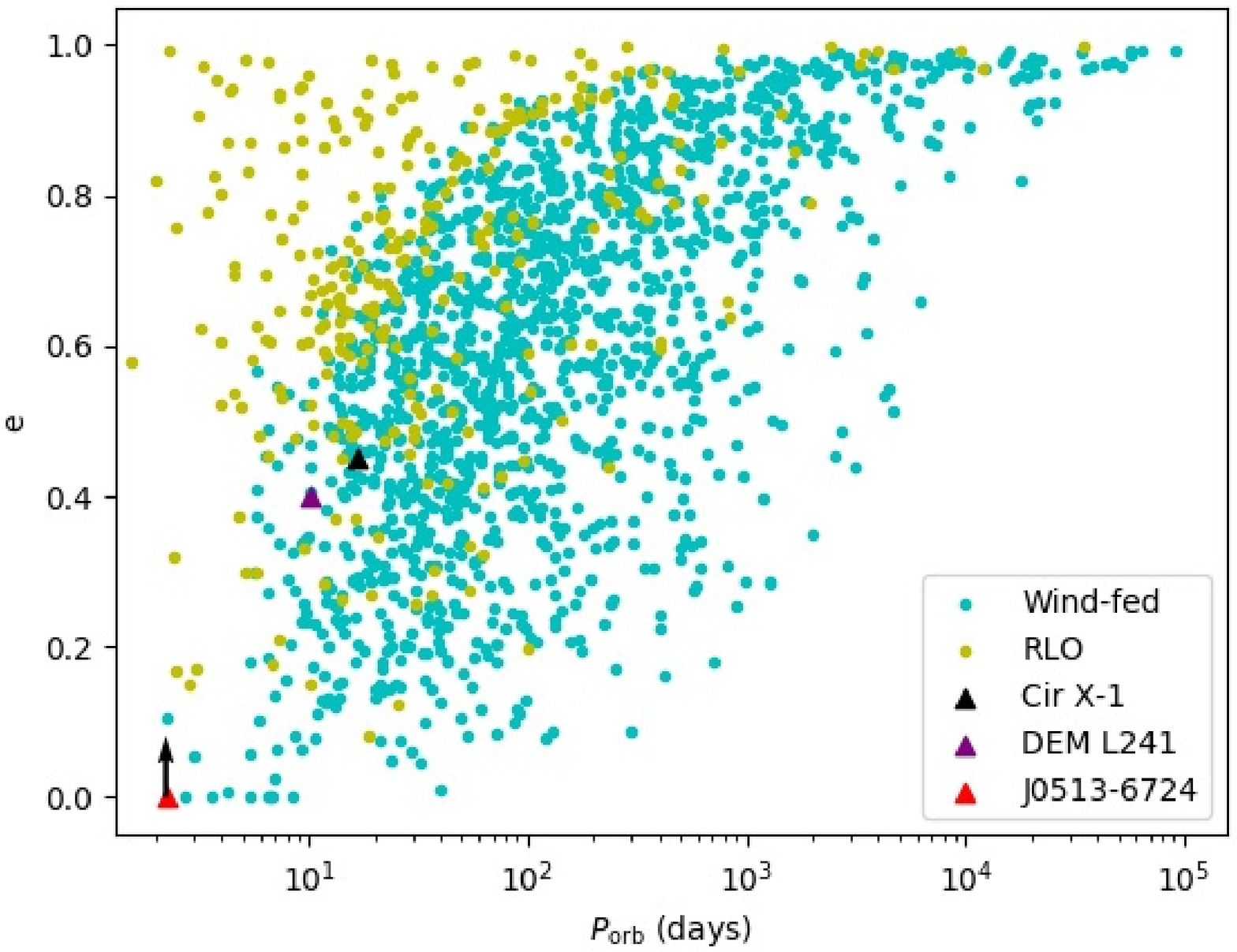}{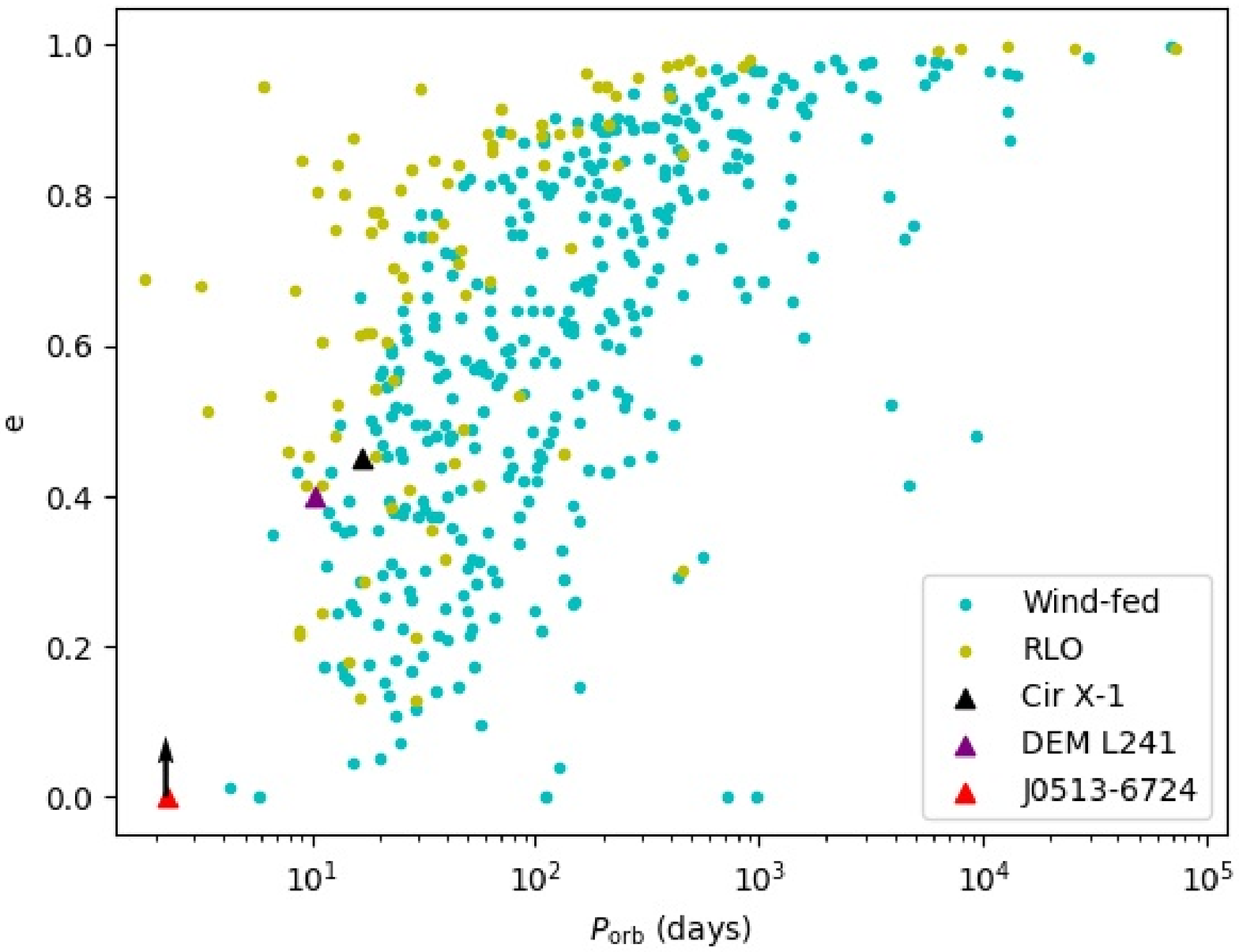}
\caption{The distributions of possible SGXBs within $10^{5}\ \rm{yr}$ after the SNe in the $P_{\rm{orb}}-e$ plane. The yellow dots represent NS-SG systems in which the donor overfills its Roche lobe at periastron and the cyan dots represent wind-fed NS-SG systems at the time they form, respectively. The black, purple, and red triangles indicate the positions of Cir X-1, LMC P3, and J0513-6724, respectively. The left and right panels correspond to $\sigma_{\rm{k,CCS}}= 150$ and $300\ \rm{km}\ \rm{s}^{-1}$, respectively.\label{fig:10}}
\end{figure}

In the calculations, we identify binaries consisting of a high-mass post-main sequence companion and a young NS. There are two possible evolutionary channels for these systems. In the first one, when the primary explodes to form a NS, the secondary has already been a supergiant, implying that the two primordial stars have very similar initial masses. In the other case, the companion star is still in its main sequence when the NS is born and leaves the main sequence within $10^{5}\ \rm{yr}$. Actually, our calculation indicates that the second channel dominates the formation of the SGXBs. In Figure \ref{fig:10} we show the orbital periods and eccentricities of the selected supergiant/NS systems that meet our requirement. The left and right panels correspond to $\sigma_{\rm{k,CCS}}= 150$ and $300\ \rm{km}\ \rm{s}^{-1}$, respectively. In each panel, the yellow dots represent the SGXBs with companion overfilling its Roche lobe at periastron when the supergiant/NS system forms and the cyan dots represent the wind-fed systems in which the components are detached at the time of the system formation. The black, purple, and red triangles depict the positions of Cir X-1, XRBs within DEM L241 and J0513-6724, respectively. In our calculations, the overall birthrates with $\sigma_{\rm{k,CCS}}= 150$ and $300\ \rm{km}\ \rm{s}^{-1}$ are $3.54\times 10^{-6}$ and $9.66\times 10^{-7}\ \rm{yr}^{-1}$, respectively. These very small birthrates raise the question why we have seen a few SGXBs associated with SNRs, which requires further studies.

\section{Discussion} \label{sec:dis}

\subsection{Supernova Ejecta-Companion Interaction}
In the above section we demonstrate the predicted properties of the companion stars at the moment of SN that produced a NS. Our results are based on the calculation of stellar evolution and mass transfer processes in binaries. However, the SNe themselves may potentially influence the structure and characteristics of the companion stars. After an SN explosion, the ejecta expands freely and impacts the companion in a few minutes to hours. The impact process can change the state of the companion by striping some mass and imparting some energy and momentum to the star. This effect, called the ejector-companion interaction (ECI), could potentially affect the results if the collision has nonnegligible influence on the binary properties of young XRBs. Some authors have investigated the ECI in terms of core-collapse supernova explosions \citep{Hi14,Hi15,Liu15,Ri16,Hi18}. They found that after suffering from mass striping, the companion is heated by the ejecta and swells up to be overluminous with lower surface temperature. The amount of the stripped mass and impact velocity are strongly tied to the explosion properties and most importantly, the binary separation. The stellar structure and how the injected energy deposited in the star will determine the long-term evolution of the companion. If the companion expands after the explosion, it may overfill its Roche lobe and transfer mass to the NS. In some extreme cases, the dramatic expansion can even make the companion engulf the NS, leading to possible mass loss or orbital circulation \citep{Hi18}.

However, after studying the interaction of CCS ejecta with low and intermediate-mass companions, \citet{Liu15} argued that the mass striping and heating effects are likely limited for most CCSNe because the separations at the moment of SNe are not small enough to cause significant changes in the companions. For massive main-sequence companions, because they have higher surface escape velocities and higher binding energy of the envelope, the effects are less significant. \citet{Hi18} studied the ECI for CCSNe in binaries with high-mass companions by numerical calculations, and paid attention to the subsequent evolution of the companions for observational implications. Particularly, they focused on a model with a $10\ M_{\odot}$ companion. Their results show that the inflated companion starts to recover after $\sim 10\ \rm{yr}$ and gets to its original state quickly within $\sim 1000\ \rm{yr}$ with the orbital separation as small as $20\ R_{\odot}$. The overall recovery timescale is about a thermal timescale of the outer layer where most injected energy is aggregated. Thus, it seems that the ECI may not strongly influence the mass-transfer state of newborn XRBs after a few $10^{3}\ \rm{yr}$ from the birth of the NS.

\subsection{The nature of Cir X-1}
The nature of the companion star in Cir X-1 has been a puzzle for years. It may be a heated main-sequence star \citep{Joh16}, a supergiant of spectral type of B5-A0 \citep{Jon07}, or a Be-like star \citep{Sch20}. \citet{Joh16} suggested that the companion of Cir X-1 may be a low-mass main-sequence star heated by the SN ejecta, which then swelled up to be a giant star and become overluminous and cool. However, as discussed above, it is quite hard for a main-sequence star to expand to fill its Roche lobe with persistent and stable mass transfer at the age of $\sim 5000\ \rm{yr}$ since its recovery timescale is quite short. The relatively small size of the star also means that the received energy from the ejecta is limited. For the representative model in \citet{Hi18}, the radius of the $10\ M_{\odot}$ main-sequence star is $5\ R_{\odot}$. If we consider the companion star to be a Sun-like star, it will only receive about $4\%$ of the energy from the ejecta comparing with a massive star with a similar separation. And there is no evidence that the SN that formed Cir X-1 was special with much higher explosion energy than the typical value of $10^{51}\ \rm{erg}$ \citep{He13}. 

On the other hand, if the companion star is a supergiant star as suggested by \citet{Jon07}, the high X-ray luminosity of the source implies that the supergiant is transferring mass via Roche-lobe overflow, because it is hard for a wind-fed system to reach such a high luminosity. With population synthesis study, our results have shown that the birthrate of such systems is pretty low, and they are too short-lived to be observed in our galaxy. Therefore, we are inclined to consider it to be a BeXRB-like system.
\section{Summary} \label{sec:sum}

We investigate the formation of young NS XRBs that are harbored in SNRs with BPS calculations and found that they have different types of donor stars. For each type of XRBs, we estimate their birthrate in Milky-Way like galaxies, analyze their overall properties, and briefly discuss their detectability. Here are the main conclusions we get:

1. In NS-main sequence Roche-lobe overfilling XRBs, the mass transfer occurs right after the formation of the NS. In most of these binaries the donor radius significantly exceeds its Roche lobe radius at periastron. And in most cases, the viscous timescale of the accretion disk is longer than the orbital period. Fewer than 10 percent of these systems are likely to have episodic mass accretion process. In the cases of $\sigma_{\rm{k,CCS}}= 150$ and $300\ \rm{km}\ \rm{s}^{-1}$, their birthrates are $9.83\times 10^{-6}$ and $5.05\times 10^{-6}\ \rm{yr}^{-1}$, respectively. 

2. BeXRBs allow a larger initial parameter space and a longer lifetime. Their birthrates with $\sigma_{\rm{k,CCS}}= 150$ and $300\ \rm{km}\ \rm{s}^{-1}$ are $1.30\times 10^{-4}$ and $4.70\times 10^{-5}\ \rm{yr}^{-1}$, respectively. Following previous studies, we assume that only the primordial stars that experienced dynamically stable mass transfer can evolve to BeXRBs. The parameter distributions we obtain are compatible with the observations.

3. For SGXBs, the birthrates from our BPS calculations are $3.54\times 10^{-6}$ and $9.66\times 10^{-7}\ \rm{yr}^{-1}$ with $\sigma_{\rm{k,CCS}}= 150$ and $300\ \rm{km}\ \rm{s}^{-1}$, respectively. These binaries originate from primordial binaries with similar-mass components, so when the primary explodes to form a NS, the secondary has become a post main-sequence star or is about to leave the main sequence within $10^{5}\ \rm{yr}$. 


\acknowledgments
This work was supported by the National Key Research and Development Program of China (2016YFA0400803), the Natural Science Foundation of China under grant Nos. 11773015, 12041301, Project U1838201 supported by NSFC and CAS, and the China Scholarship Council (CSC).

%

\vspace{5mm}

\end{document}